\documentclass[12pt,a4paper]{article}
\usepackage{graphicx}
\usepackage{amsmath,amssymb,epsfig}
\usepackage{longtable}
\usepackage{cite}
\usepackage{url}
\usepackage{multirow}
\usepackage{changes}
\usepackage{rotating}
\usepackage{tabularx}
\usepackage{verbatim}

\newcommand{\aspi}{\frac{\hat\alpha_s}{\pi}}
\newcommand{\etal}{{\em et al.}}
\newcommand{\be}{\begin{equation}}
\newcommand{\ee}{\end{equation}}
\newcommand{\ba}{\begin{array}}
\newcommand{\ea}{\end{array}}
\newcommand{\bea}{\begin{eqnarray}}
\newcommand{\eea}{\end{eqnarray}}
\usepackage{color}

\begin{document}

\title{
{\hfill \normalsize MITP/22-012} \\ 
\vspace*{24pt}
\Large\bf Bottom Quark Mass with Calibrated Uncertainty}
\author{ \large Jens Erler$^\star$ and Hubert Spiesberger$^\parallel$ \\ 
{\normalsize \em PRISMA$^+$ Cluster of Excellence, Institute for Nuclear Physics$^\star$ and Institute of Physics$^\parallel$,} \\
{\normalsize \em Johannes Gutenberg-University, 55099 Mainz, Germany} \\ \\
Pere Masjuan \\
{\normalsize \em Grup de F\'{\i}sica Te\`orica, Departament de F\'{\i}sica, Universitat Aut\`onoma de Barcelona,} \\
{\normalsize \em and Institut de F\'{\i}sica d'Altes Energies, Barcelona Institute of Science and Technology,} \\
{\normalsize \em Campus UAB, E-08193 Bellaterra (Barcelona), Spain}}

\maketitle

\begin{abstract}
We determine the bottom quark mass $\hat{m}_b$ from QCD sum rules of moments of the vector current correlator calculated 
in perturbative QCD to ${\cal O} (\hat\alpha_s^3)$. 
Our approach is based on the mutual consistency across a set of moments where experimental data are required 
for the resonance contributions only. 
Additional experimental information from the continuum region can then be used for stability tests and to assess the theoretical uncertainty. 
We find $\hat{m}_b(\hat{m}_b) = (4180.2 \pm 7.9)$~MeV for $\hat\alpha_s(M_Z) = 0.1182$. 
\end{abstract}


\section{Introduction}
Highly precise values of the charm and bottom quark masses can be obtained in QCD perturbation theory, 
because they are sufficiently large to suppress non-perturbative effects. 
The object of interest is the vector current correlation function, 
which can be studied experimentally in a clean environment in electron-positron annihilation. 
Furthermore, by considering moments of the correlator one arrives at theoretically most accessible inclusive observables, which 
--- at least in the case of the vector current --- offers excellent perturbative convergence even in the context of the charm quark mass, $m_c$, 
where the strong coupling, $\alpha_s(m_c) \sim 0.4$, is not all that small. 
By the specific method~\cite{Erler:2002bu} reviewed in Sect.~\ref{sec:moments}, 
which is a concrete implementation of the general QCD sum rule idea~\cite{Shifman:1978bx,Shifman:1978by}, 
we were able to determine $m_c$ with a controlled theory uncertainty~\cite{Erler:2016atg}, 
competitive even with the results from lattice gauge theory simulations~\cite{Komijani:2020kst}, 
and in excellent agreement with them~\cite{Zyla:2020zbs}.

In the case of the bottom quark mass, $m_b$, lattice gauge theory faces an impediment, as the strong interaction scale of 
${\cal O}(m_\rho)$ differs significantly from $m_b$ itself.
By contrast, this separation of scales is a virtue in any approach effectively utilizing the operator product expansion (OPE).
Together with the smaller value of the strong coupling, $\alpha_s(m_b) \sim 0.23$, this turns the QCD sum rule approach 
into the method of choice to determine $m_b$.

On the other hand, much less experimental information is available on the bottom quark current correlator compared to that of the charm quark. 
This is because the $b\bar b$ electro-production cross section is by more than an order of magnitude smaller than 
non-$b\bar b$ quark production, so that $B$ tagging is needed in order to determine the exclusive cross section. 
Furthermore, while formally the domain of the dispersion integration extends to infinite energy, 
the experimentally scanned kinematical region for bottom meson pair production does not exceed $\sqrt{s} \approx 11.2$~GeV, 
leaving a roughly four times smaller window in relative comparison to open charm production. 
Fortunately, this problem can be solved by considering higher moments, which in contrast to the charm 
case~\cite{Erler:2016atg,Chetyrkin:2010ic,Bodenstein:2011ma,Dehnadi:2015fra} is a viable option for $m_b$.

The essential feature of our approach (Sect.~\ref{sec:moments}) is that the masses and electronic decay widths 
of the low-lying $\Upsilon$ resonances provide sufficient experimental knowledge to determine $m_b$, 
as long as the 0$^{th}$ moment is considered alongside the more standard positive-$n$ moments. 
We may then use the limited experimental information from the continuum region that {\em is\/} available to test the stability of our results
in Sect.~\ref{sec:results} as a function of the moment number, and to control (in fact over-constrain) the theoretical uncertainty 
(see Sect.~\ref{sec:data}). 
We present our conclusions and a comparison with other approaches in Sect.~\ref{sec:conclusions}.

\section{Moment sum rules}
\label{sec:moments}
The transverse part of the correlation function $\hat\Pi_q(t)$ 
(quantities marked with a caret are defined in the $\overline{\rm MS}$ renormalization scheme) 
of two heavy quark vector currents obeys the subtracted dispersion relation~\cite{Chetyrkin:1994js}, 
\be
12 \pi^2 \frac{\hat\Pi_q (0) - \hat\Pi_q (-t)}{t} = \int\limits_{4 \hat m_q^2}^\infty \frac{{\rm d} s}{s} \frac{R_q(s)}{s + t}\ ,
\label{eq:SR}
\ee
where $R_q(s) = 12 \pi \mbox{Im} \hat\Pi_q(s)$, and where $\hat m_q = \hat m_q (\hat m_q)$ is the heavy quark mass.
Taking derivatives in the limit $t \rightarrow 0$, one obtains the moments~\cite{Shifman:1978bx,Shifman:1978by,Novikov:1977dq},
\be
{\cal M}_n := \left.\frac{12\pi^2}{n !} \frac{d^n}{d t^n} \hat\Pi_q(t) \right|_{t=0} = \int\limits_{4 \hat m_q^2}^\infty \frac{{\rm d} s}{s^{n+1}} R_q(s) 
\qquad\qquad (n \geq 1).
\label{eq:SRder}
\ee
A $0^{th}$ moment~\cite{Erler:2002bu} can also be defined, 
\be
{\cal M}_0 := - \lim_{t \to \infty} \left[ \hat\Pi(-t) - \hat\Pi^\infty(-t) \right] = 
\int\limits_{\hat m_q^2}^\infty \frac{{\rm d} s}{s} \left[ R_q(s) - R_q^\infty(s) \right],
\label{eq:SRder0}
\ee
provided the limit $t \to \infty$ and the integration over $\mbox{Im} \hat\Pi(s)$ at $s \to \infty$ is regularized by properly 
chosen subtractions $\hat\Pi^\infty(-t)$ and $R_q^\infty(s)$ \cite{Erler:2002bu,Erler:2016atg}. ${\cal M}_0$ is then obtained 
from the dispersion relation for the difference $\hat\Pi_q (-t) - \hat\Pi_q (0)$ where the (unphysical) constant $\hat\Pi_q (0)$ is subtracted. 
The subtraction and the explicit sum rule for ${\cal M}_0$ will be given below. 

The left-hand sides of Eqs.~(\ref{eq:SRder}) and (\ref{eq:SRder0}) can be calculated in perturbative QCD (pQCD) order by order 
in the strong coupling $\hat\alpha_s(\hat m_q)$ as a function of $\hat m_q$. 
On the right-hand side one can use the optical theorem to relate $R_{q}(s)$ to the cross section for heavy quark production 
in $e^+e^-$ annihilation. 
It can be split into a contribution from a small number of narrow resonances below the heavy quark production 
threshold and a continuum contribution above,
\be
R_q(s) = R_q^{\rm res}(s) + R_q^{\rm cont}(s).
\label{eq:Rq}
\ee 

One possible method to determine $\hat m_q$ is thus to combine data, where available, for the evaluation of the integrals on 
the right-hand side of Eqs.~(\ref{eq:SRder}) and (\ref{eq:SRder0}) with predictions from pQCD at large $s$ where there are no data. 
This approach has been followed, for example, in Refs.~\cite{Chetyrkin:2010ic,Dehnadi:2015fra,Kuhn:2001dm,Kuhn:2007vp,Chetyrkin:2009fv}. 
A certain amount of modeling is necessary since experimental information about $R_q(s)$ is restricted to relatively small energies. 

Here, we will choose a different strategy. 
The idea is to describe the continuum region above the heavy quark production threshold on average only, 
not having to rely on local quark-hadron duality. 
We follow Refs.~\cite{Erler:2002bu,Erler:2016atg} and use the simple {\em ansatz\/},
\begin{eqnarray} 
R_q^{\rm cont}(s) = 3 Q^2_q \lambda^q_1 (s) \sqrt{1 - \frac{4\, \hat{m}_q^2 (2 M)}{s^\prime}} 
\left[1 + \lambda^q_3 \left( \frac{2\, \hat{m}_q^2(2 M)}{s^\prime} \right) \right],
\label{eq:ansatz}
\end{eqnarray}
where $3 Q_q^2 \lambda_1^q(s)$ is the zero-mass limit of $R_{q}(s)$ and $s' := s + 4 [\hat{m}_q^2(2M) - M^2]$. $M$ is 
taken as the mass of the lightest pseudoscalar meson, {\em i.e.}\ in the case of the bottom quark, $M = M_{B^{\pm}} = 5.27934$~GeV~\cite{Zyla:2020zbs}. 
This {\em ansatz\/} guarantees a smooth transition between the onset of the heavy quark production threshold at $2 M$ and pQCD at large $s$. 
Since we only need to consider moments, fine details of the {\em ansatz\/} are not very important. 
However, we will also investigate variations of our {\em ansatz\/} where the resonances above the threshold $4M^2$, 
$\Upsilon(4S)$, $\Upsilon(5S)$, and $\Upsilon(6S)$, are explicitly added to the expression~(\ref{eq:ansatz}).

\begin{table}[t]
\begin{center}
\begin{tabular}{|r|l|r|l|l|}
\hline \rule[2pt]{0pt}{13pt}
$n$ & $~~C_n^{(0)}$ & $C_n^{(1)}~~$ & $~~C_n^{(2)}$ & $~~~~~C_n^{(3)}$ \\[2pt] 
\hline \rule[1pt]{0pt}{13pt}
$1$ & $1.0667$ & $2.5547$ & $3.1590$ & $-7.7624$  \\[1pt]
$2$ & $0.4571$ & $1.1096$ & $3.2320$ & $-2.6438$ \\[1pt]
$3$ & $0.2709$ & $0.5194$ & $2.0677$ & $-1.1745$ \\[1pt]
$4$ & $0.1847$ & $0.2031$ & $1.2205$ & $-1.60 \pm 0.5$ \\[1pt]
$5$ & $0.1364$ & $0.0106$ & $0.7023$ & $-2.29 \pm 1.2$ \\[1pt]
$6$ & $0.1061$ & $-0.1159$ & $0.4304$ & $-2.73 \pm 1.8$ \\[1pt]
$7$ & $0.0856$ & $-0.2033$ & $0.3358$ & $-2.85 \pm 2.3$ \\[1pt]
$8$ & $0.0709$ & $-0.2660$ & $0.3701$ & $-2.80 \pm 2.7$ \\[1pt]
$9$ & $0.0601$ & $-0.3122$ & $0.4988$ & $-2.75 \pm 3.1$ \\[1pt]
$10$ &$0.0517$ & $-0.3470$ & $0.6979$ & $-2.68 \pm 3.4$ \\[1pt] 
\hline
\end{tabular}
\end{center}
\caption{Coefficients $C_n^{(i)}$ for the perturbative expansion of the QCD moments entering Eq.~(\ref{eq:Cnth}). 
The values quoted with an uncertainty are taken from Ref.~\cite{Greynat:2011zp}.}
\label{tab:Cni}
\end{table}

The two unknowns, namely the heavy quark mass $\hat{m}_q(\hat{m}_q)$, and the single free parameter in Eq.~(\ref{eq:ansatz}), 
$\lambda^q_3$, will be determined from Eq.~(\ref{eq:SRder0}) and one of the Eqs.~(\ref{eq:SRder}). 
The other moments are then fixed and can be used to check the consistency of the approach~\cite{Erler:2016atg}. 
Thus, besides the value of $M$, only the masses and electronic decay widths of the low-lying resonances are needed 
as the experimental input to extract $\hat{m}_q(\hat{m}_q)$. 
The quark mass and $\lambda_3^q$ can, in principle, be determined from any combination of two moments. 
But only including the $0^{th}$ moment provides the leverage to sufficiently break the correlation between $\lambda_3^q$ from $\hat{m}_q$.

We now give the explicit expressions needed for our numerical evaluation. 
From now on we particularize to the bottom quark case, in which we may neglect higher-dimensional operators in the OPE, 
such as from the gluon condensate\footnote{A positive gluon condensate reduces $\hat{m}_b(\hat{m}_b)$ by at most $0.2$~MeV 
with only mild moment dependence, which is well below other uncertainties.}.
Perturbative QCD predictions for the positive moments can be cast into the form,
\be
{\cal M}_n^{\rm pQCD} = \frac{1}{4} \left( \frac{1}{2\hat{m}_b(\hat{m}_b)} \right)^{2n} \hat{C}_n\ ,
\label{eq:Mth}
\ee
with
\be
\hat{C}_n = C_n^{(0)} + \left(\aspi + \frac{\alpha_{\rm em}}{12 \pi}\right) C_n^{(1)} + \frac{\hat\alpha_s^2}{\pi^2} C_n^{(2)} + 
\frac{\hat\alpha_s^3}{\pi^3} C_n^{(3)} + {\cal O}(\hat\alpha_s^4).
\label{eq:Cnth}
\ee
The coefficients $\hat{C}_n$ are known~\cite{Chetyrkin:2006xg,Boughezal:2006px,Kniehl:2006bf,Maier:2008he,Maier:2009fz} up to 
${\cal O}(\hat\alpha_s^3)$ for $n \leq 3$, and up to ${\cal O}(\hat\alpha_s^2)$ for the rest~\cite{Chetyrkin:1997mb,Maier:2007yn}. 
The numerical values required for our analysis are collected in Table~\ref{tab:Cni}. 

\begin{figure}[t]
\begin{center}
\includegraphics[width=0.8\textwidth]{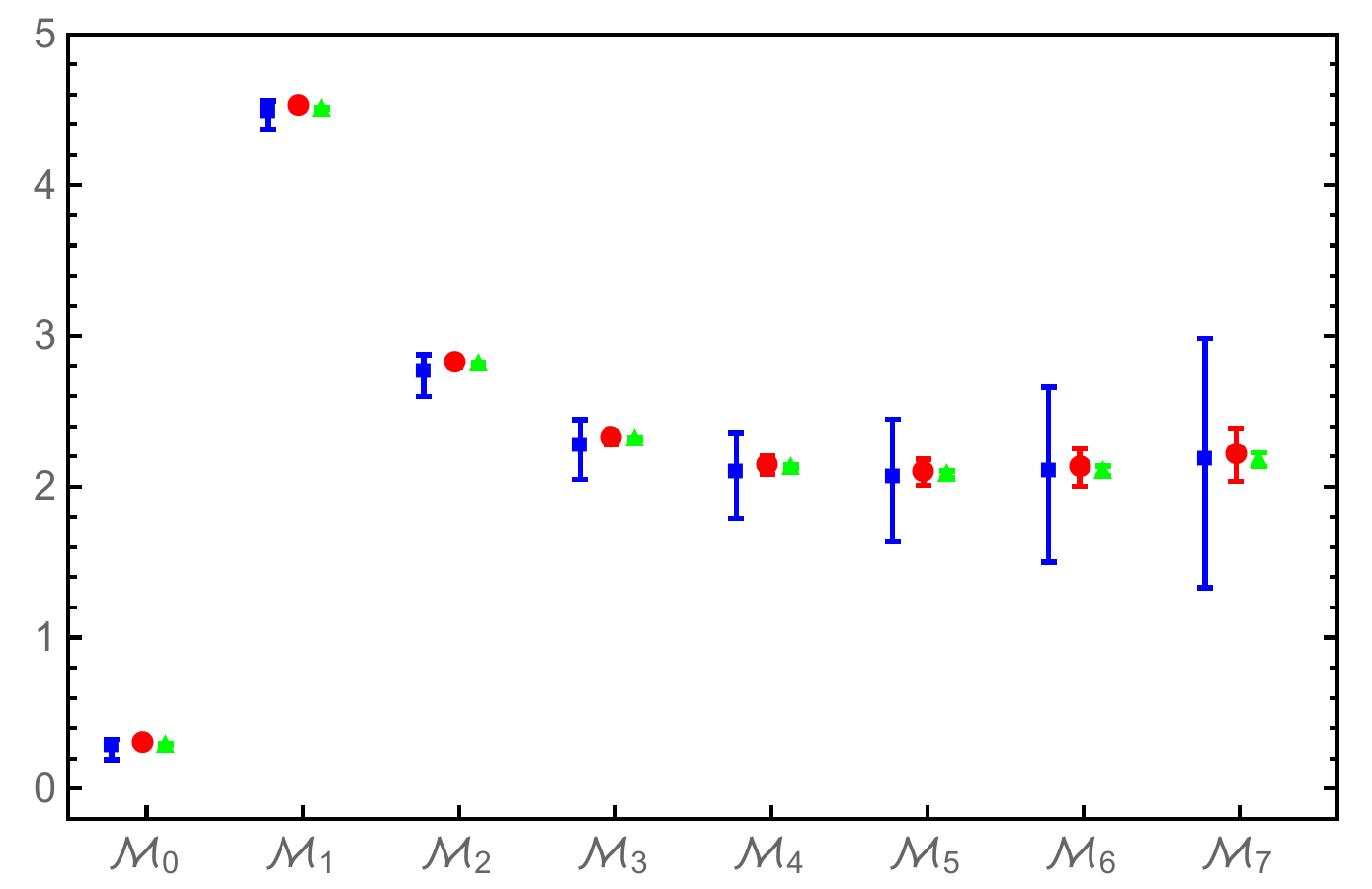}
\caption{The theoretical moments ${\cal M}_n^{\rm pQCD}$ in Eq.~(\ref{eq:Mth}) (multiplied by $10^{2n+1} \mbox{ GeV}^{4n+2}$) 
for the reference value $\hat{m}_b(\hat{m}_b) = 4.180$~GeV at different orders in $\hat\alpha_s$. 
Blue squares show results at ${\cal O}(\hat\alpha_s)$, red circles include ${\cal O}(\hat\alpha_s^2)$ terms, 
and green triangles refer to the full ${\cal O}(\hat\alpha_s^3)$. 
The error bars are the truncation uncertainties from Eq.~(\ref{eq:truncerror}) at the given order.}
\label{fig:bmmoments}
\end{center}
\end{figure}

Since we evaluate the moments up to ${\cal O} (\hat\alpha_s^3)$, we use the predictions for $n>3$ provided in Ref.~\cite{Greynat:2011zp}, 
inducing the uncertainties shown in the table. 
In the approach of Ref.~\cite{Greynat:2011zp}, based on the Mellin-Barnes transform, the two-point correlator at ${\cal O}(\alpha_s^3)$ 
is reconstructed from the Taylor expansion at $q^2=0$, the threshold expansion at $q^2 = 4\hat m_q^2$, 
and the high-energy expansion at $q^2 \to \infty$. 
The reconstruction is analytic and systematic, and is controlled by an error function which becomes smaller as more terms in the expansions 
are known. 
Once the correlator is reconstructed, one can calculate the moments in Eq.~(\ref{eq:SRder}). 
An overview of our theory errors for the moments up to $n=7$ is shown in Fig.~\ref{fig:bmmoments}. 
An alternative prediction of these coefficients in Ref.~\cite{Kiyo:2009gb} has quoted errors that are smaller by one order of magnitude or more, 
leading to a total error in the extracted $\hat m_b$ about an MeV smaller, but we believe that the conservative approach 
of Ref.~\cite{Greynat:2011zp} is a better reflection of the corresponding error.

We shall approximate the contributions of the narrow resonances by $\delta$-functions, 
\be
R_b^{\rm res}(s) = \sum_{R = \Upsilon(1S), \Upsilon(2S), \Upsilon(3S)} \frac{9\pi}{\alpha_{\rm em}^2(M_R)} M_R \Gamma_R^e \delta(s - M_R^2),
\label{eq:Rres}
\ee
where the masses $M_R$ and electronic widths $\Gamma^e_R$~\cite{Zyla:2020zbs} are listed in Table~\ref{tab:ResPDG}. 
The values of the running fine structure constant at the resonance are also given in the table\footnote{The values for $\alpha_{\rm em}(M_R)$ 
were determined with the help of the program {\tt hadr5n12}~\cite{jegerlehner-url}.}. 

\begin{table}[t]
\begin{center}
\begin{tabular}{|r|l|c|l|c|}
\hline \rule[2pt]{0pt}{13pt}
$R$~~~ & $M_R$ [GeV] & $\Gamma_R$ & $\Gamma_R^e$ [keV] & $\alpha^2_{\rm em}(0)/\alpha^2_{\rm em}(M_R)$ \\[2pt] 
\hline \rule[1pt]{0pt}{13pt}
$\Upsilon(1S)$ & \phantom{0}9.46030 & 54.02(1.25) keV & 1.340(18) & 0.931308 \\[1pt]
$\Upsilon(2S)$ & 10.02326 & 31.98(2.63) keV & 0.612(11) & 0.930113 \\[1pt]
$\Upsilon(3S)$ & 10.3552 & 20.32(1.85) keV & 0.443(8) &  0.929450 \\[1pt] 
\hline \rule[1pt]{0pt}{13pt}
$\Upsilon(4S)$ & 10.5794 & 20.5(2.5) MeV & 0.272(29) & 0.929009 \\[1pt]
$\Upsilon(5S)$ & 10.8852 & ~~37~(4)~~~MeV & 0.31(7) & 0.928415 \\[1pt]
$\Upsilon(6S)$ & 11.000 & ~~24~(7)~~~MeV & 0.130(30) & 0.928195 \\[1pt] 
\hline
\end{tabular}
\end{center}
\caption{Resonance data~\cite{Zyla:2020zbs} used in the analysis. 
The uncertainties from the resonance masses are negligible. 
The first three resonances are below the continuum threshold and define $R_b^{\rm res}(s)$, 
while the higher ones will be needed later when we evaluate the theoretical uncertainties.}
\label{tab:ResPDG}
\end{table}

Finally, we need the regularized expression for ${\cal M}_0$, which requires to subtract the zero-mass limit of 
$R_{b}(s) = 3 Q_b^2 \lambda_1^b(s)$. 
While it is known to order ${\cal O}(\hat\alpha_s^4)$, we need only the third-order expression~\cite{Chetyrkin:2000zk},
\begin{align}
\lambda^b_1(s) &= 1 + \frac{\hat \alpha_s}{\pi} + \frac{3 Q_b^2 \alpha_{\rm em}}{4 \pi} \left(1-\frac13  \frac{\hat \alpha_s}{\pi} \right) 
+ \frac{\hat \alpha_s^2}{\pi^2} \left[ \frac{365}{24} - 11 \zeta(3) + n_b \left( \frac{2}{3} \zeta(3) - \frac{11}{12} \right) \right] \nonumber \\[6pt]
&+ \frac{\hat \alpha_s^3}{\pi^3} \left[\frac{87029}{288} - \frac{121}{8} \zeta(2) - \frac{1103}{4} \zeta(3) + \frac{275}{6}\zeta(5) \right.
\label{eq:lambda1} \\[6pt]
&+ \left. n_b \left(- \frac{7847}{216} + \frac{11}{6} \zeta(2) + \frac{262}{9} \zeta(3) - \frac{25}{9}\zeta(5) \right) + 
n_b^2 \left(\frac{151}{162} - \frac{\zeta(2)}{18} - \frac{19}{27} \zeta(3) \right) \right], \nonumber 
\end{align}
where $\hat \alpha_s = \hat \alpha_s (\sqrt s)$,  $\alpha_{\rm em} = \alpha_{\rm em}(\sqrt s)$, and $n_b = 5$ is the total number of active flavors. 
Using the results of Refs.~\cite{Chetyrkin:1996cf,Chetyrkin:1997un}, the sum rule for ${\cal M}_0$ defined in Eq.~(\ref{eq:SRder0}) reads explicitly,
\begin{align} 
&\sum\limits_{\rm resonances} \frac{27\pi\Gamma^e_R}{M_R \alpha_{\rm em}^2 (M_R)} + 
\int\limits_{4 M^2}^\infty \frac{{\rm d} s}{s} \left[ 3 R_b^{\rm cont}(s) - \lambda_1^b(s) \right] - 
\int\limits_{\hat{m}_b^2}^{4 M^2} \frac{{\rm d} s}{s} \lambda^b_1 (s) = \nonumber \\[6pt]
&- \frac{5}{3} + \aspi \left[ 4 \zeta(3) - \frac{7}{2} \right] + \frac{\hat\alpha_s^2}{\pi^2} 
\left[ \frac{2429}{48} \zeta(3) - \frac{25}{3} \zeta(5) - \frac{2543}{48} + n_b \left( \frac{677}{216} - \frac{19}{9} \zeta(3) \right) \right] + 
\frac{\hat\alpha_s^3}{\pi^3} A_3 \nonumber \\[6pt]
&= - 1.667  + 1.308\, \aspi + 2.192 \frac{\hat\alpha_s^2}{\pi^2} - 8.117 \frac{\hat\alpha_s^3}{\pi^3}\ ,
\label{eq:SR0}
\end{align}
where $\hat\alpha_s = \hat \alpha_s(\hat m_b)$. 
The third-order coefficient $A_3$ is available in numerical form~\cite{Erler:2016atg,Hoang:2008qy,Kiyo:2009gb},
\be
A_3 = -9.863 + 0.399 \, n_b - 0.010 \, n_b^2\ .
\ee
In the last line of Eq.~(\ref{eq:SR0}) we show the numerical values for $n_b=5$. 
The onset of the continuum is at $2M$, the pseudoscalar threshold. 
The lower integration limit in the subtraction term involving $\lambda_1^b(s)$ is, in principle, arbitrary, but is set to $\hat{m}_b^2$ 
in concordance with the choice to evaluate $\hat{\alpha}_s$ on the right-hand side of Eq.~(\ref{eq:SR0}) at scale $\hat{m_b}$. 

For both the theoretical predictions of the moments and the contributions from resonances and continuum to the sum rules
one has to assess the uncertainties. 
To assign a truncation error to the pQCD prediction of the moments we follow the method proposed in Refs.~\cite{Erler:2002bu,Erler:2016atg} 
and consider the largest group theoretical factor in the next un-calculated perturbative order, 
\be
\Delta {\cal M}_n^{(i)} = \pm Q_q^2 N_C C_F C_A^{i-1} \left[ \frac{\hat\alpha_s (\hat{m}_q)}{\pi}\right]^i 
\left[ \frac{1}{2 \hat{m}_q(\hat{m}_q)} \right]^{2n},
\label{eq:truncerror}
\ee
($N_C = C_A = 3$, $C_F = 4/3$). 
Alternatively, the dependence on the renormalization scale is often used to estimate theory errors, where, for example, 
in Refs.~\cite{Kuhn:2007vp,Dehnadi:2015fra} the scale was varied between $5$ and $15$~GeV. 
Our prescription is more conservative, as has already been observed in our previous analysis~\cite{Erler:2016atg} of the charm quark mass. 

In order to determine the error from the continuum contribution we proceed as follows. 
First, we choose a pair of moments $({\cal M}_0, {\cal M}_n)$ from which $\hat{m}_b(\hat{m}_b)$ and $\lambda_3^{b}$ are determined. 
Then we input this value of $\hat{m}_b(\hat{m}_b)$ into Eq.~(\ref{eq:ansatz}) and integrate with the weight corresponding to the $0^{th}$ moment 
as in Eq.~(\ref{eq:SRder0}), but with the energy integration range restricted to the threshold region, 
$2 M_B\leq \sqrt{s} \leq 11.20~\mbox{GeV}$. 
As this is a function of $\lambda_3^b$, we can adjust its value to coincide with the corresponding integral over the experimentally determined 
threshold region (see Sect.~\ref{sec:data}) yielding an {\em experimental\/} value, denoted $\lambda_3^{b,{\rm exp}}$. 
In the final step, we use $\lambda_3^{b,{\rm exp}}$ in the $n^{th}$ moment sum rule to re-calculate $\hat{m}_b(\hat{m}_b)$, 
and treat the difference between these two $\hat{m}_b(\hat{m}_b)$ values as an additional uncertainty. 
It serves as a control of the error component associated with the entire methodology 
which we will denote by $\lambda_3^b \neq \lambda_3^{b,{\rm exp}}$. 
For example, neglected non-perturbative contributions to the moments such as from condensates or from residual duality violations 
would become visible in the comparisons of the values $\lambda_3^b$ from the theoretical moments with $\lambda_3^{b,{\rm exp}}$.
The experimental errors in the threshold data induce an uncertainty $\Delta\lambda_3^{b,{\rm exp}}$ in $\lambda_3^{b,{\rm exp}}$ itself, 
which we will also need to account for.

\section{Numerical results and determination of $\hat{m}_b$}
\label{sec:results}
We have analyzed the determination of $\hat{m}_b(\hat{m}_b)$ from different pairs of moments and using different prescriptions 
to include resonances on top of the continuum. 
The results are shown in figures and tables in this section. 
We find that the largest source of uncertainty is from the continuum contribution. 
Indeed, the values of $\lambda_3^b$ derived from the mutual consistency of the moments deviate from $\lambda_3^{\rm exp}$ 
determined from data if none of the resonances above threshold are taken into account explicitly. 
The lower moments are more sensitive to the continuum region, and this deviation indicates that the simple {\em ansatz\/} 
using only $R_q^{\rm cont}(s)$ from Eq.~(\ref{eq:ansatz}) does not capture the strong onset of the cross section for energies 
just above the threshold for open bottom production. 
As a consequence, stable results are not reached for lower moments. 
However, the stability improves greatly with the inclusion of the $\Upsilon(4S)$ and $\Upsilon(5S)$ states. 
We parametrize them as Gamma distributions,
\begin{equation}
R_b^{\rm res, Gamma}(s) = 
\sum_{R = \Upsilon(4S),\Upsilon(5S)} \frac{9\pi}{\alpha_{\rm em}^2(M_R)} \Gamma_R^eM_R {\rm Gamma}(s - 4 M_B^2| \alpha, \beta),
\label{eq:Gamma} 
\end{equation}
where 
\begin{equation}
{\rm Gamma}(x|\alpha, \beta) := \frac{\beta^\alpha}{\Gamma(\alpha)} x^{\alpha-1} e^{-\beta x}, \hspace{60pt} (\alpha > 0,\ \beta > 0),
\label{eq:GammaDistribution} 
\end{equation}
and $\alpha$ and $\beta$ are chosen such that the peak location $M_R$ and the second derivative coincide with those 
of a relativistic Breit-Wigner distribution with width $\tilde\Gamma_R$,
\begin{equation}
\alpha = 1 + \frac{2}{\sqrt[3]{\pi}} \frac{(M_R^2 - 4 M_B^2)^2}{\tilde\Gamma_R^2 M_R^2}\ , \hspace{60pt} 
\beta = \frac{\alpha - 1}{M_R^2 - 4 M_B^2}\ . 
\label{eq:Gammapars}
\end{equation}
We use the peak positions $M_R$ and total width $\Gamma_R$ of the resonances as given in Ref.~\cite{Zyla:2020zbs} and collected above in  Table~\ref{tab:ResPDG}. 

\begin{figure}[t!]
\begin{center}
\includegraphics[width=0.46\textwidth]{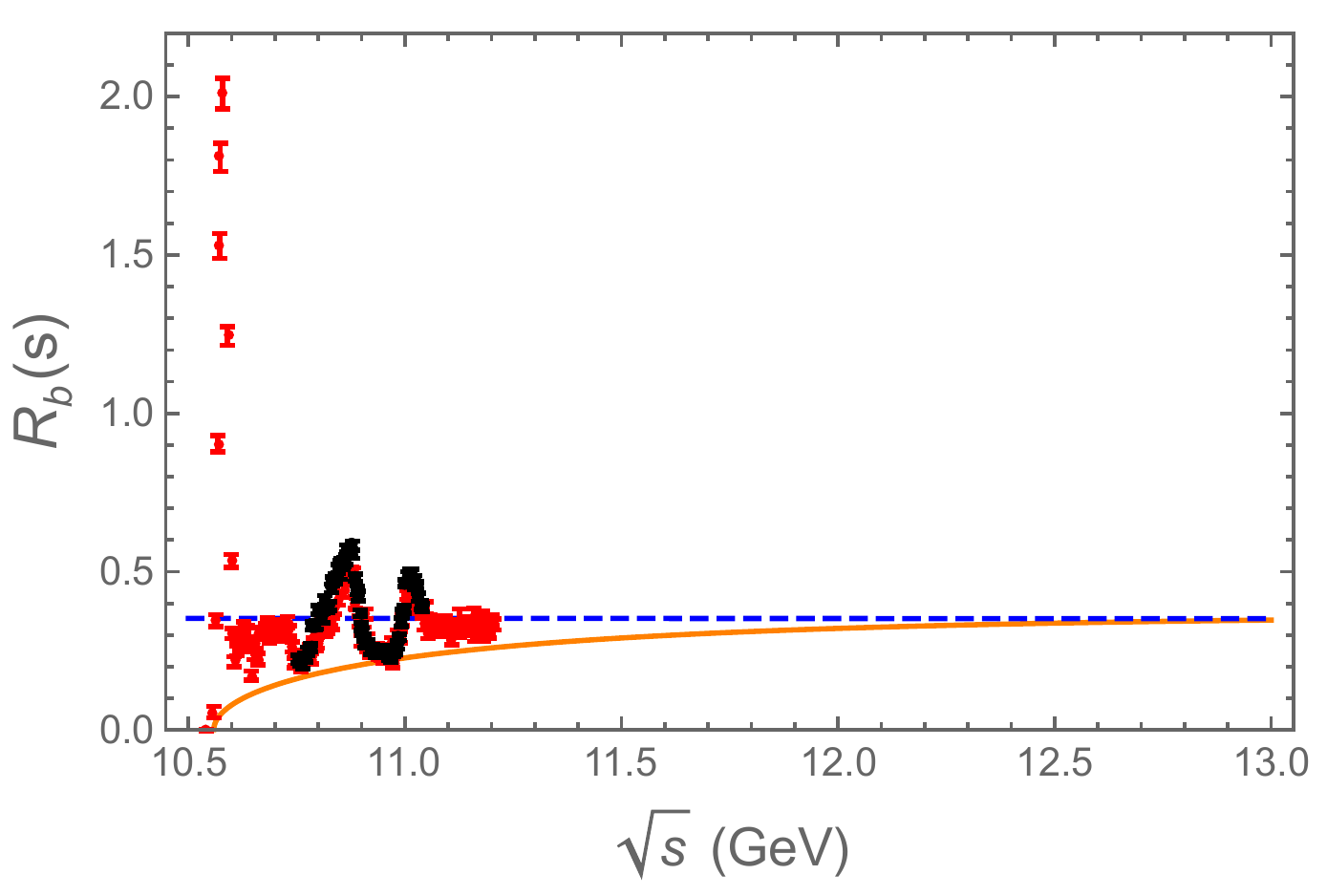} \\
\includegraphics[width=0.46\textwidth]{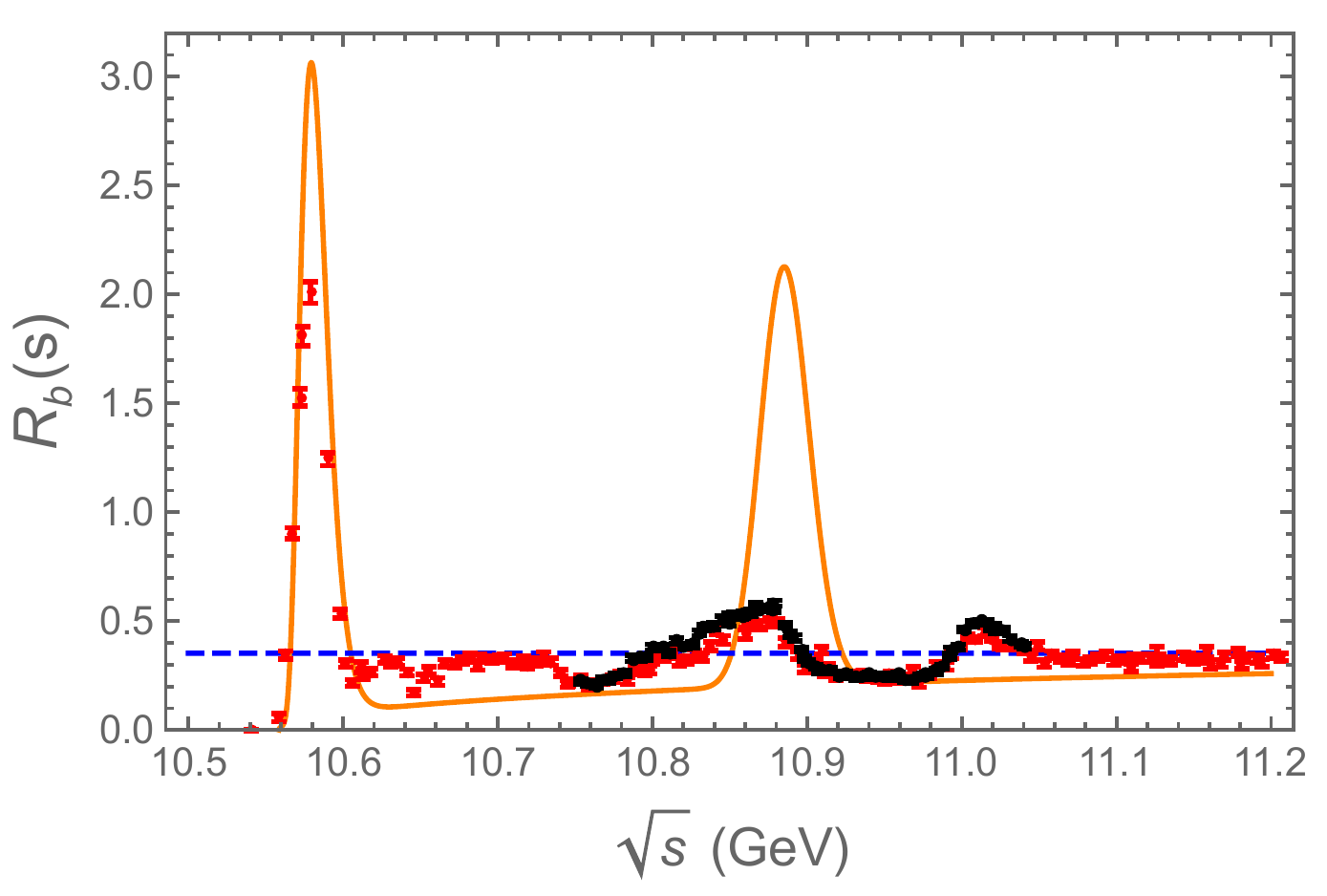} 
\includegraphics[width=0.46\textwidth]{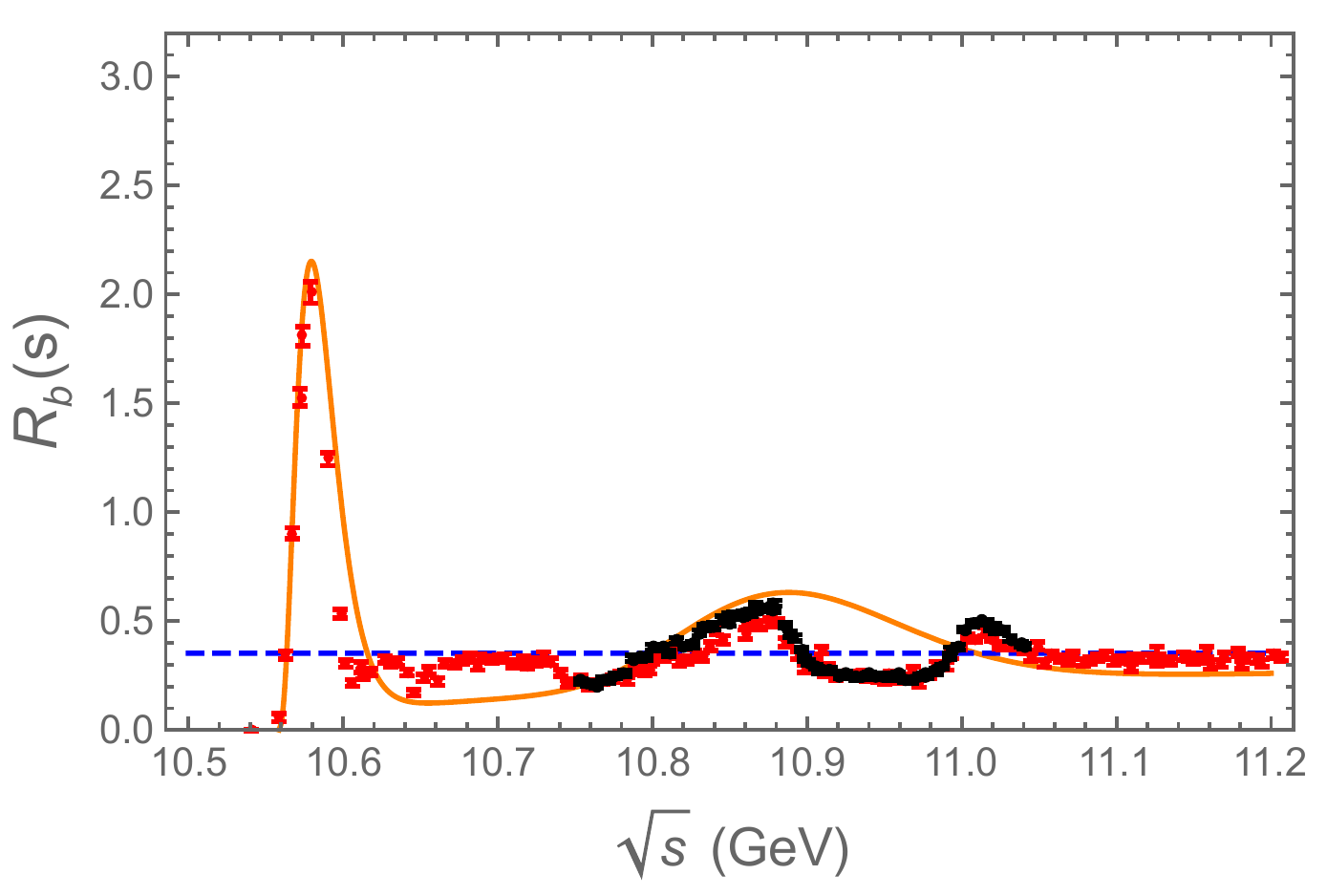} \\
\includegraphics[width=0.46\textwidth]{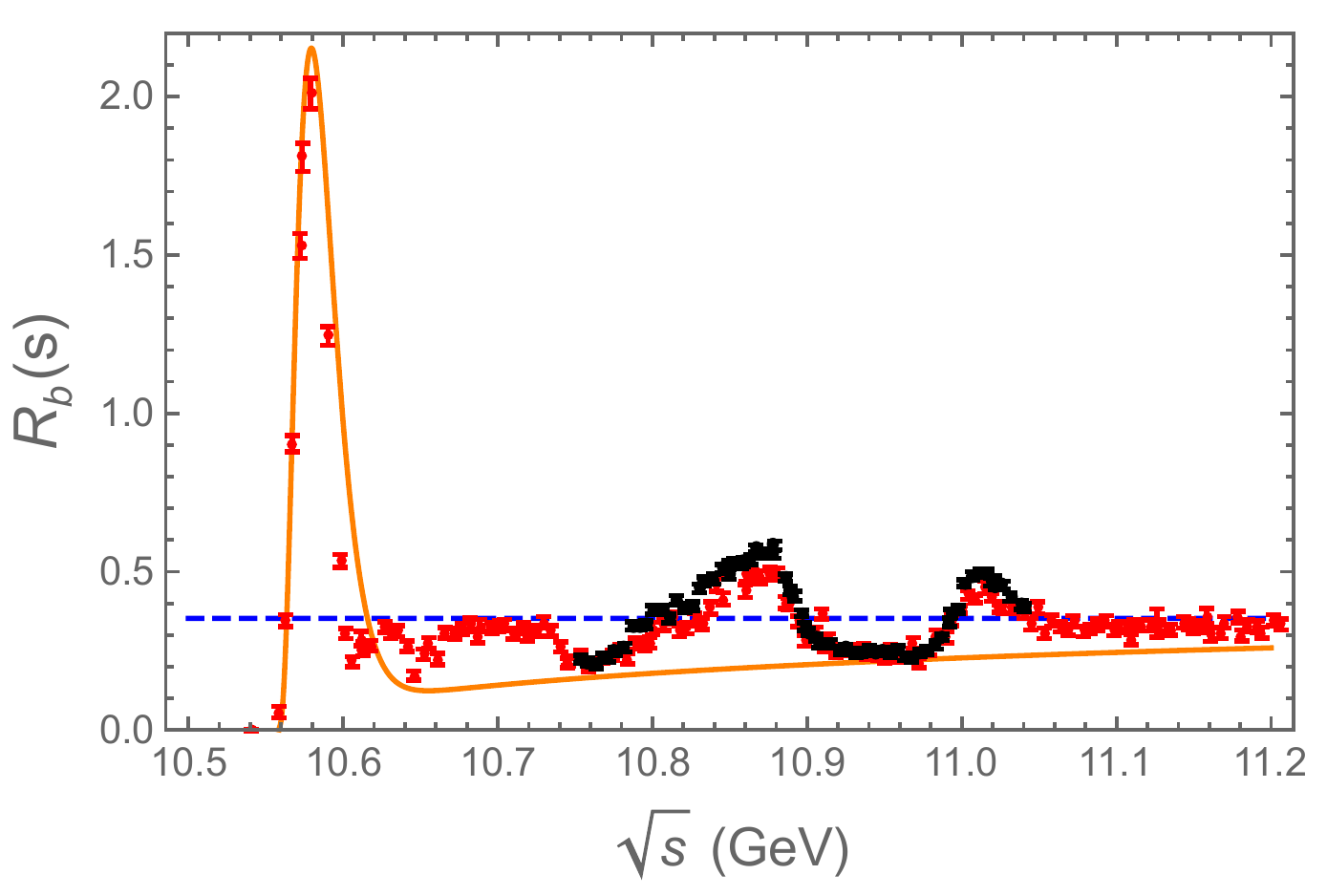} 
\includegraphics[width=0.46\textwidth]{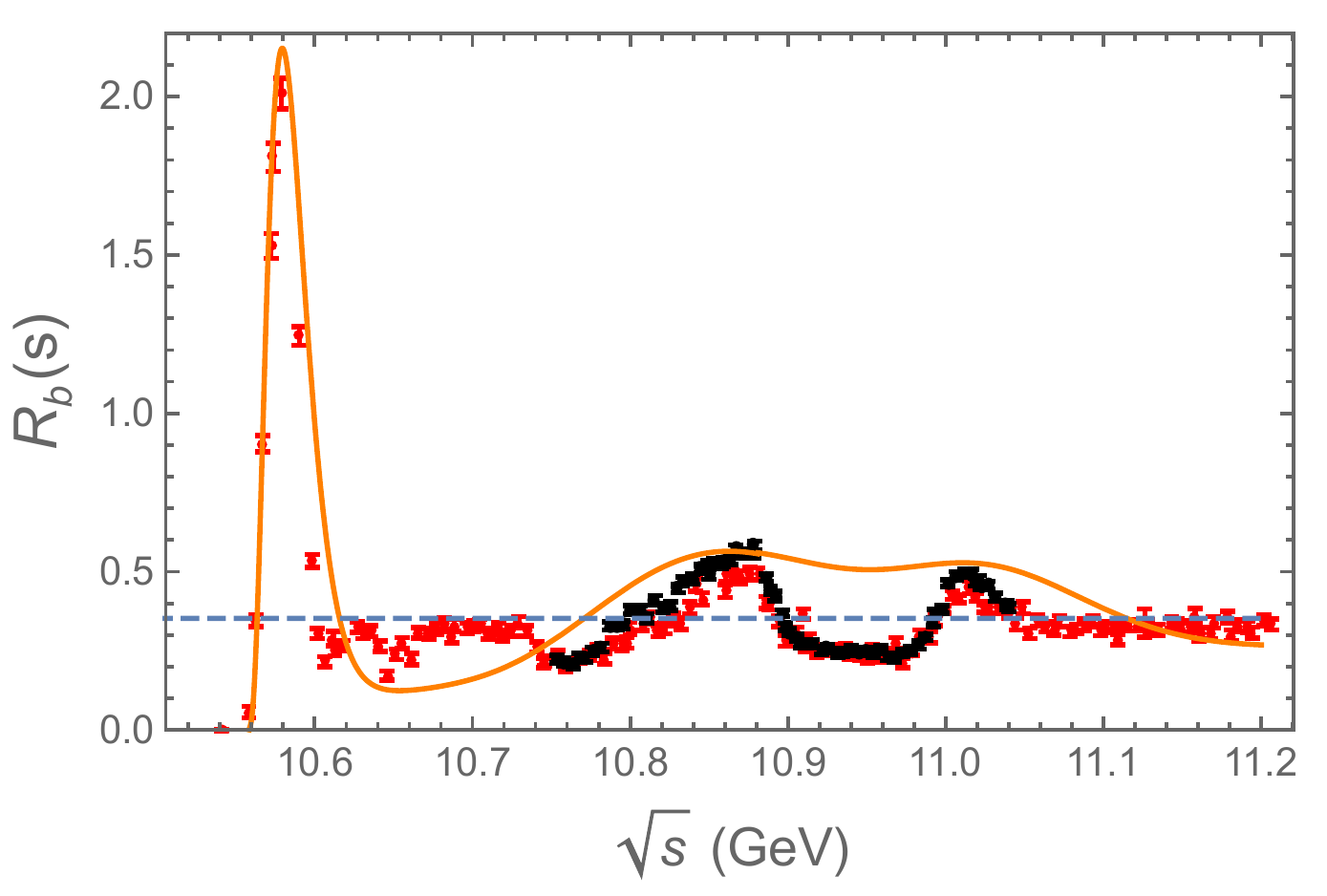} 
\caption{$R_b(s)$ from ISR corrected  BaBar\cite{Aubert:2008ab} data (red points) and from Belle\cite{Santel:2015qga} data (black points) 
compared with different choices for our {\em ansatz\/} for continuum plus resonances. 
The blue dashed line is the pQCD prediction for $R_b(s)$. 
Upper plot: Continuum {\em ansatz\/} without resonances extended up $\sqrt{s} = 13$~GeV above the range where data are available. 
Middle row: Continuum {\em ansatz\/} including Gamma distributions for the $\Upsilon(4S)$ and $\Upsilon(5S)$ resonances. 
In the left plot (our default choice) widths $\Gamma_R$ from the PDG data are used (see Table~\ref{tab:ResPDG}), 
while in the right plot the widths are tuned to match the local description of the data, 
$\tilde\Gamma_{\Upsilon(4S)} = 29$~MeV and $\tilde\Gamma_{\Upsilon(5S)} = 165$~MeV. 
Lower row: Alternative choices including only the $\Upsilon(4S)$ resonance on top of the continuum 
(left, $\tilde\Gamma_{\Upsilon(4S)} = 29$~MeV), or the three resonances $\Upsilon(4S)$, $\Upsilon(5S)$, and $\Upsilon(6S)$
(right, $\tilde\Gamma_{\Upsilon(4S)} = 29$~MeV, $\tilde\Gamma_{\Upsilon(5S)} = 192$~MeV, $\tilde\Gamma_{\Upsilon(6S)} = 139$~MeV).}
\label{fig:RbvsA}
\end{center}
\end{figure}

\begin{table}[t]
\begin{center}
\begin{tabular}{|l|rrrrrr|}
\hline \rule[2pt]{0pt}{13pt}
& $({\cal M}_0, {\cal M}_1)$ 
& $({\cal M}_0, {\cal M}_3)$ 
& $({\cal M}_0, {\cal M}_5)$ 
& $({\cal M}_0, {\cal M}_6)$ 
& $({\cal M}_0, {\cal M}_7)$ 
& $({\cal M}_0, {\cal M}_8)$ \\[2pt] 
\hline \rule[1pt]{0pt}{13pt}
$\hat{m}_b(\hat{m}_b)$ & 4224.0 & 4187.4 & 4181.1 & 4180.4 & 4180.2 & 4180.1 \\[1pt] \rule[0pt]{0pt}{13pt}
$\lambda_3^b$ & 1.45 & 1.52 & 1.53 & 1.53 & 1.53 &  1.53 \\[1pt] \rule[0pt]{0pt}{13pt}
$\lambda_3^{b,{\rm exp}}$ & 0.85(20) & 0.83(20) & 0.82(20) & 0.82(20) & 0.82(20) & 0.82(20) \\[1pt] 
\hline\hline \rule[1pt]{0pt}{13pt}
Resonances & 12.3 & 6.8 & 4.3 & 3.7 & 3.2 & 2.8 \\[1pt] \rule[0pt]{0pt}{13pt}
Truncation & 2.9 & 2.9 & 4.1 & 5.0 & 6.3& 7.9 \\[1pt] \rule[0pt]{0pt}{13pt}
$\lambda_3^b \neq \lambda_3^{b,{\rm exp}}$ & $-113.7$ & $-28.5$ & $-9.3$ & $-5.6$ & $-3.5$ & $-2.2$ \\[1pt] \rule[0pt]{0pt}{13pt}
$\Delta\lambda_3^{b,{\rm exp}}$ & $39.0$ & $8.2$ & $2.6$ & $1.6$ & $1.0$ & $0.6$ \\[1pt] \hline \rule[1pt]{0pt}{13pt}
Total & $120.9$ & $30.6$ & $11.3$ & $8.5$ & $7.9$ & $8.7$ \\[1pt] \hline \rule[1pt]{0pt}{13pt}
$10^3\Delta \hat\alpha_s(M_Z)$ 
& $-5.54 \Delta \hat\alpha_s$ 
& $-0.93 \Delta \hat\alpha_s$ 
& $-0.28 \Delta \hat\alpha_s$ 
& $-0.18 \Delta \hat\alpha_s$ 
& $-0.11 \Delta \hat\alpha_s$ 
& $-0.07 \Delta \hat\alpha_s$ \\[1pt] \rule[0pt]{0pt}{13pt}
EW fit & $\mp8.9$ & $\mp1.5$ & $\mp0.5$ & $\mp0.3$ & $\mp0.2$ & $\mp0.1$ \\[1pt] \hline
\end{tabular}
\end{center}
\caption{Above the double line: values of $\hat{m}_b(\hat{m}_b)$ (in MeV), $\lambda_3^b$ and $\lambda_3^{b,{\rm exp}}$, 
determined from different pairs of moments as described in the text, where the $\Upsilon(4S)$ and $\Upsilon(5S)$ resonances 
have been added explicitly to the {\em ansatz\/} in Eq~(\ref{eq:ansatz}). 
Below the double line: breakdown of the uncertainties in $\hat{m}_b(\hat{m}_b)$ followed by the total errors. 
The dependence on $\hat\alpha_s$ is shown in the next-to-last line, where the minus sign indicates that $\hat{m}_b$ decreases 
when $\hat\alpha_s$ is increased relative to the reference value $\hat\alpha_s(M_Z) = 0.1182$. 
The last line contains the uncertainty induced from $\Delta \hat\alpha_s(M_Z) = \pm 0.0016$, {\em i.e.}\ the error obtained from the global fit 
to electroweak precision data~\cite{Zyla:2020zbs}. 
See also Fig.~\ref{fig:bmass4S5S} for a graphical representation of these results.}
\label{tab:MomentsBudget4S5S}
\end{table}

\begin{figure}[t!]
\begin{center}
\includegraphics[width=1.\textwidth]{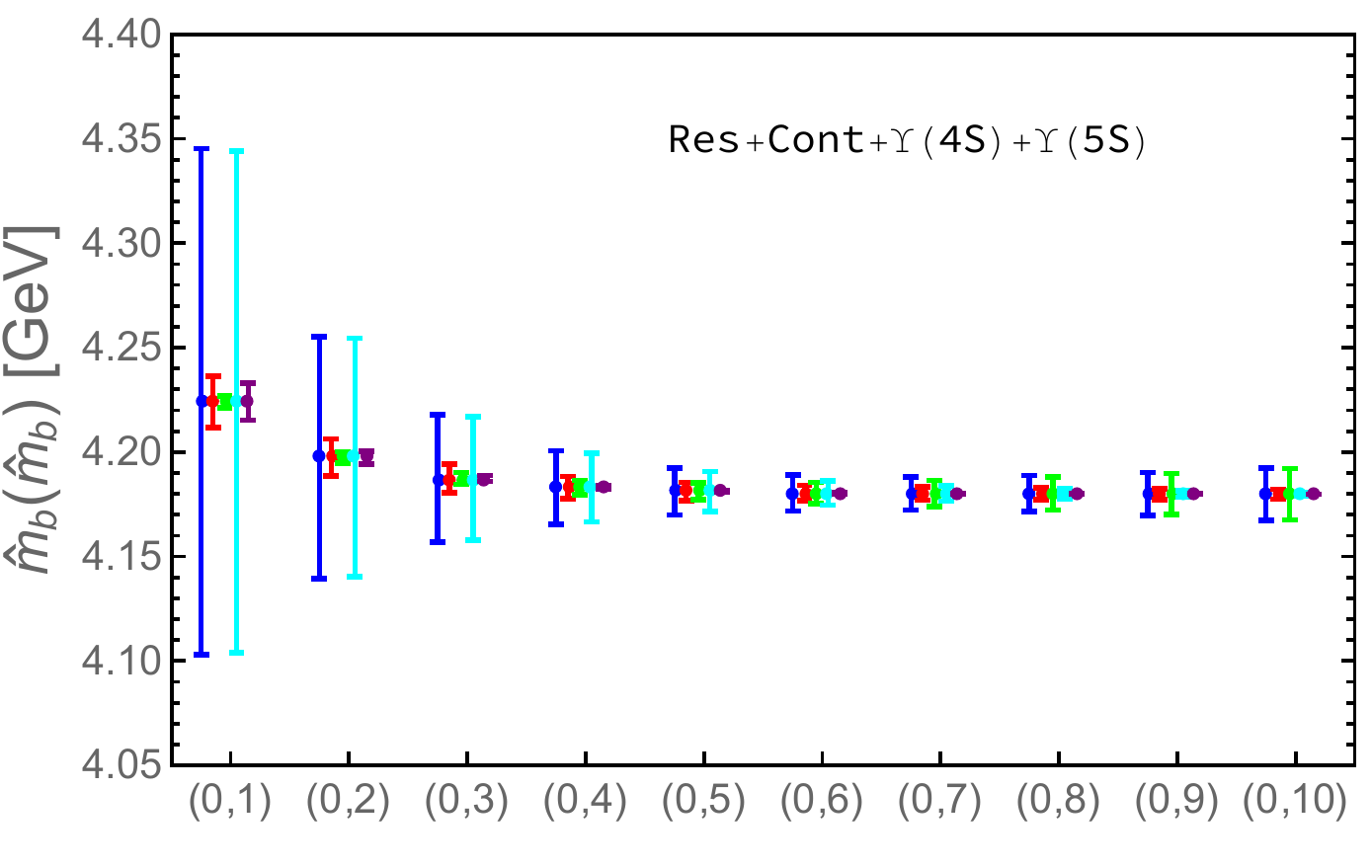}
\caption{Results for $\hat{m}_b(\hat{m}_b)$ using different combinations of moments, 
where we added the $\Upsilon(4S)$ and $\Upsilon(5S)$ states explicitly to the {\em ansatz\/} in Eq.~(\ref{eq:ansatz}), as described in the text. 
Blue bars represent the full error, 
red bars are from the experimental uncertainties in the resonance parameters, 
green bars indicate the truncation errors in the theoretical moments, 
cyan bars are the symmetrized error combinations due to $\lambda_3^b \neq \lambda_3^{b,{\rm exp}}$ and $\Delta\lambda_3^b$ 
(see Table~\ref{tab:MomentsBudget4S5S}), 
and the uncertainty induced by $\Delta \hat\alpha_s(M_Z) = \pm 0.0016$ is shown in purple.}
\label{fig:bmass4S5S}
\end{center}
\end{figure}

In order to understand why the inclusion of resonances above threshold lead to an improved determination of the $b$ quark mass, 
we provide in Fig.~\ref{fig:RbvsA} a graphical account of the landscape of $R_b(s)$ above threshold. 
The upper plot shows that the continuum alone does not describe the data in the energy range of the $\Upsilon(4S)$, $\Upsilon(5S)$, 
and $\Upsilon(6S)$ resonances and the pQCD limit is reached only when $\sqrt{s}$ is above threshold by an amount 
of the order of the $b$ quark mass, {\em i.e.}\ far above the energy range where data are available. 
It is therefore not a suprise that with the continuum {\em ansatz\/} alone one cannot obtain stable solutions from the set of sum rules. 
The second row of plots in Fig.~\ref{fig:RbvsA} shows how the global description of data for $R_b(s)$ 
can be improved by the inclusion of Gamma distributions, Eq.~(\ref{eq:Gamma}), for the $\Upsilon(4S)$ and $\Upsilon(5S)$ resonances. 
If we use the total decay widths $\Gamma_R$ in Eq.~(\ref{eq:Gammapars}) as given by the PDG~\cite{Zyla:2020zbs} 
the local description of the data is still not good; however, the moments, {\em i.e.}\ integrals over $R_b(s)$ can be matched. 
To see this more clearly one can exploit the fact that moments do not change  even if the total widths are significantly increased (which we denote by $\tilde\Gamma_R$) if one aims at a better visual representation of the local behavior of the data, as done for the right plot of the middle row of Fig.~\ref{fig:RbvsA}. 
Here a good description of the data {\em on average} is clearly visible. 
The lower row of plots in Fig.~\ref{fig:RbvsA} shows other possible choices, namely to add only one resonance, 
the $\Upsilon(4S)$, or three resonances, $\Upsilon(4S)$, $\Upsilon(5S)$ and $\Upsilon(6S)$, on top of the continuum. 
The first (latter) choice would lead to an underestimate (overestimate) of moments in the region above threshold. 
As a consequence, these choices would lead to solutions for $\lambda_3^b$ from the set of sum rules in disagreement 
with $\lambda_3^{b,{\rm exp}}$ as determined from data. 

\begin{figure}[t!]
\begin{center}
\includegraphics[width=0.64\textwidth]{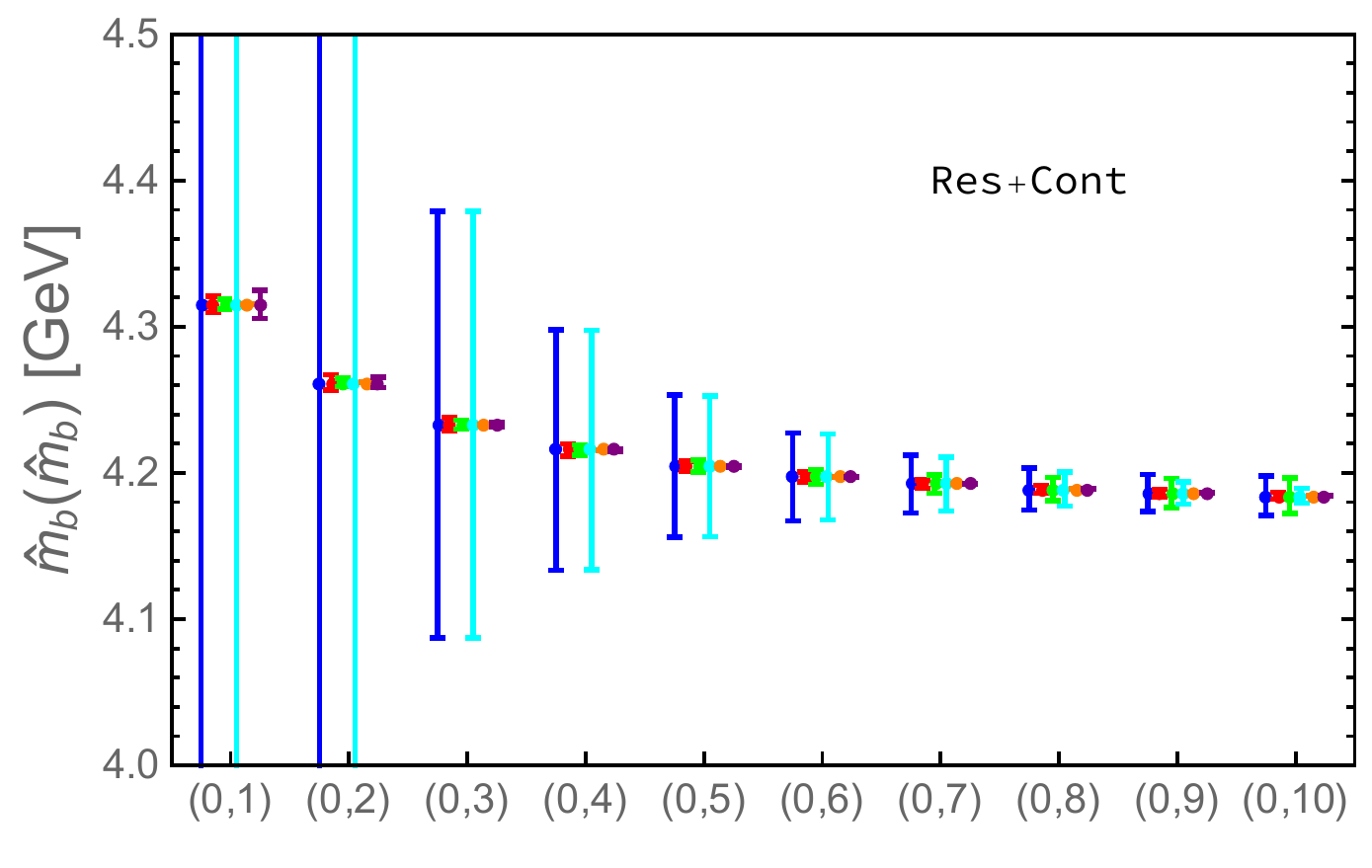}
\includegraphics[width=0.64\textwidth]{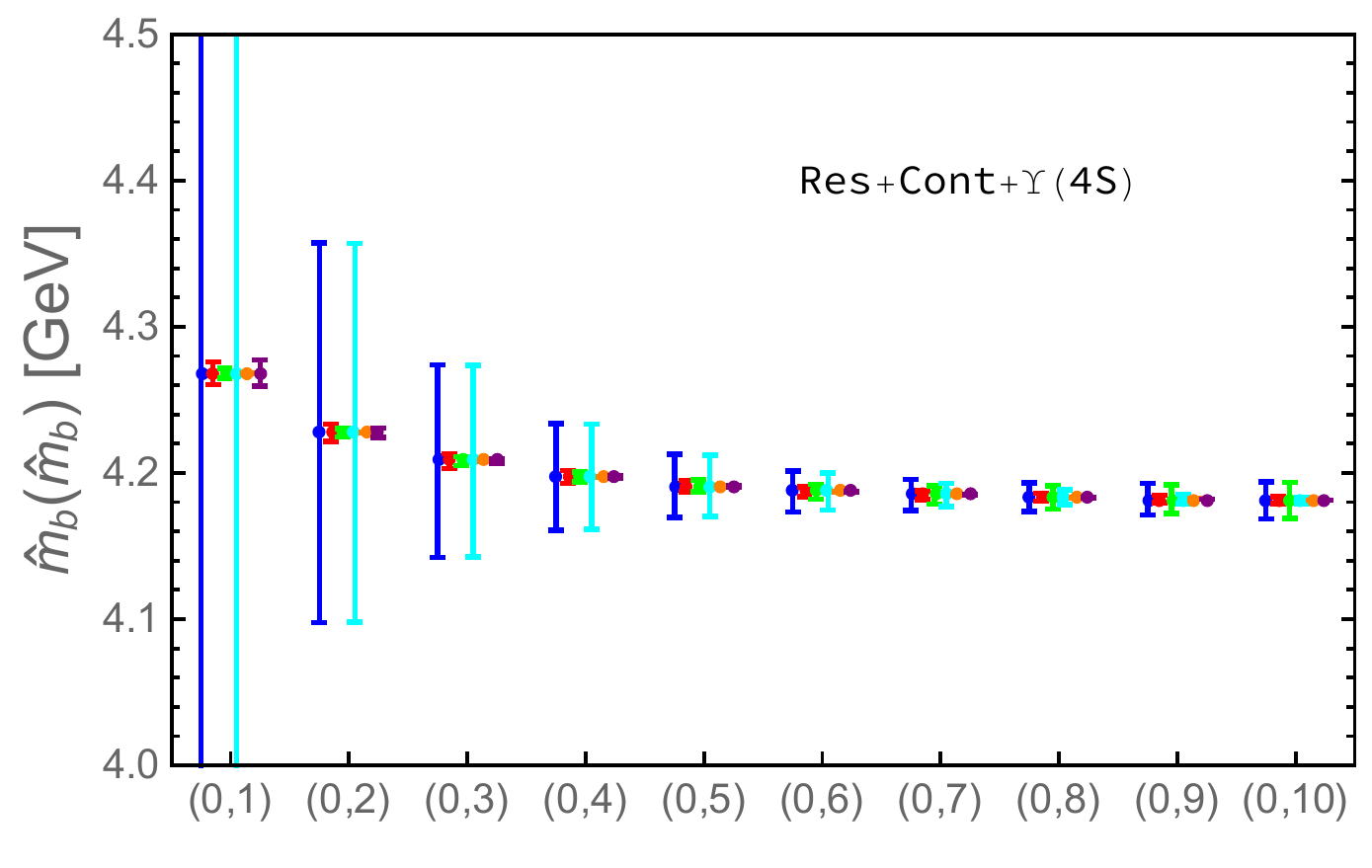}
\includegraphics[width=0.64\textwidth]{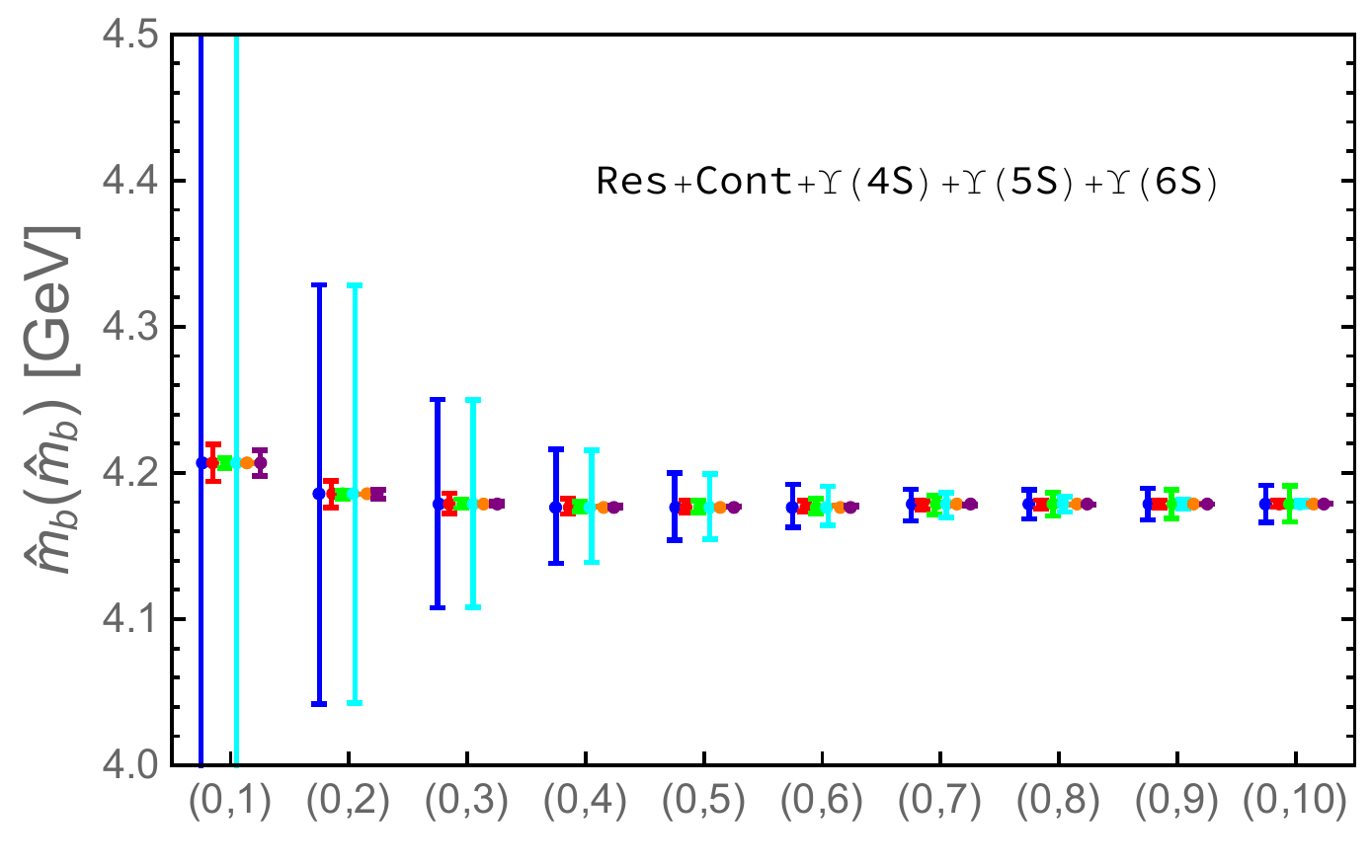}
\caption{Same as in Fig.~\ref{fig:bmass4S5S} but including only the resonances below threshold (top), 
with the {\em ansatz\/} modified to include the $\Upsilon(4S)$ as a $\Gamma$-function (middle), 
and with the {\em ansatz\/} modified to include the $\Upsilon(4S)$, $\Upsilon(5S)$ and $\Upsilon(6S)$ as $\Gamma$-functions (bottom).}
\label{fig:bmass}
\end{center}
\end{figure}

We therefore determine the two free parameters, $\lambda_3^b$ and $\hat{m}_b(\hat{m}_b)$ from pairs of sum rules using 
the continuum {\em ansatz\/} where we include the $\Upsilon(4S)$ and $\Upsilon(5S)$ as described above. 
The results are summarized in Table~\ref{tab:MomentsBudget4S5S} and Fig.~\ref{fig:bmass4S5S}, including the breakdown of the uncertainties 
from the different sources as discussed before. 
Results for other options, 
(i) where we do not include resonances on top of the continuum, 
(ii) where we include only the $\Upsilon(4S)$, 
(iii) or where we include additionally the $\Upsilon(6S)$ parametrized as a Gamma distribution as well, are presented in Fig.~\ref{fig:bmass}. 
The shift of the value for the $b$ quark mass induced by these different options is small; 
for example including three resonances above threshold, $\hat{m}_b(\hat{m}_b)$ would be reduced by 1.3~MeV, 
{\em i.e.}\ by much less than our error estimate. 
The most stable result and smallest overall uncertainty is obtained with our default option in Fig.~\ref{fig:bmass4S5S}.

A summary of the best determination in each scenario is shown in Table~\ref{tab:MomentsBudgetSummary}. 
Our most precise and therefore final result for $\hat m_b(\hat m_b)$ is based on the pair of moments $({\cal M}_0, {\cal M}_7)$, and reads,
\begin{equation}
\hat m_b(\hat m_b) = (4180.2 - 108.5 \Delta \hat{\alpha_s} \pm 7.9)~{\rm MeV}.
\label{eq:bmresult}
\end{equation} 
We explicitly exhibit the dependence on the input value of the strong coupling $\hat \alpha_s$ relative to the central value\footnote{This 
value corresponds to $\hat\alpha_s(\hat m_b(\hat m_b)) = 0.225$, 
where we have used five-loop running~\cite{Baikov:2016tgj,Herzog:2017ohr} with four-loop matching~\cite{Schroder:2005hy} of $\hat\alpha_s$.}, 
{\em i.e.}\ $\Delta \hat \alpha_s = \hat \alpha_s(M_Z) - 0.1182$. 

\begin{table}[t]
\begin{center}
\begin{tabular}{|l|c|c|}
\hline \rule[2pt]{0pt}{13pt}
& $\hat{m}_b(\hat{m}_b)$ [MeV] & Pair of moments \\[2pt] 
\hline \rule[1pt]{0pt}{13pt}
Only resonances below threshold & $ 4186.7 - ~39.5\, \Delta \hat{\alpha}_s \pm 12.7 $ & $({\cal M}_0, {\cal M}_9)$ \\[1pt]
+ $\Upsilon(4S)$ & $4183.8 - ~68.0\; \Delta \hat{\alpha}_s \pm ~9.7$ & $({\cal M}_0, {\cal M}_8)$ \\[1pt]
+ $\Upsilon(4S)+\Upsilon(5S)$ & $ 4180.2 - 108.5\, \Delta \hat{\alpha}_s \pm ~7.9 $ & $({\cal M}_0, {\cal M}_7)$ \\[1pt]
+ $\Upsilon(4S)+\Upsilon(5S)+\Upsilon(6S)$ & $ 4178.9 - ~64.0\; \Delta \hat{\alpha}_s \pm ~9.7 $ & $({\cal M}_0, {\cal M}_8)$ \\[1pt] \hline
\end{tabular}
\end{center}
\caption{Values and uncertainties of the bottom quark mass when adding various resonances on top of the continuum {\em ansatz\/}. 
Only the values with the smallest uncertainty and the corresponding pair of moments from which it is obtained are shown in each case.}
\label{tab:MomentsBudgetSummary}
\end{table}

\section{Experimental moments}
\label{sec:data}
Our determination of $\hat m_b(\hat m_b)$ described above does not rely on the details of experimental data for $R_b$ 
except resonance parameters. 
However, a comparison with data for $R_b$ allows us to calibrate the uncertainty of the $\hat m_b(\hat m_b)$ determination. 
As described above, this is done by calculating moments from data and extracting an experimental value for $\lambda^{b,{\rm exp}}_3$ 
which can be compared with the value of $\lambda^b_3$ obtained from the consistency relations for moments. 
In this section we present the details of our determination of $\lambda^{b,{\rm exp}}_3$. 

We take data from the  BaBar Collaboration~\cite{Aubert:2008ab}. 
These data cover the range of energies between $\sqrt{s} = 10.54$ and 11.20~GeV ({\em cf.}\ Fig.~\ref{RawBabarData}). 
Data from the Belle Collaboration~\cite{Santel:2015qga} will be used to obtain a cross-check, 
but they cover too short a range in energies to be useful for a calculation of moments for our purpose.

\subsection{Data and corrections}
The published experimental data for continuum heavy quark production must be corrected for vacuum polarisation and QED radiative effects 
before they can be used in our analysis. 
Corrections due to vacuum polarisation can be taken into account by substituting the value for $\alpha_{\rm em}$ used in the experimental work 
by the running fine structure constant, $\alpha_{\rm em}(\sqrt{s})$. 
Since the variation of $\alpha_{\rm em}(\sqrt{s})$ in the considered energy range is very small, 
we take it to be constant and use $(\alpha_{\rm em}(0)/\alpha_{\rm em}(M_R))^2 = 0.93$, (see Table~\ref{tab:ResPDG}). 
This factor should be multiplied with the measured $R_b$ ratio. 

 BaBar experimental data are available for energies above the open bottom threshold. 
In this energy range, the radiative tails from the $\Upsilon(1S)$, $\Upsilon(2S)$, $\Upsilon(3S)$ resonances contribute. 
The required corrections are provided by  BaBar in supplementary material to Ref.~\cite{Aubert:2008ab} and are easily subtracted from the data. 

\begin{figure}[t!]
\begin{center}
\includegraphics[width=0.75\linewidth]{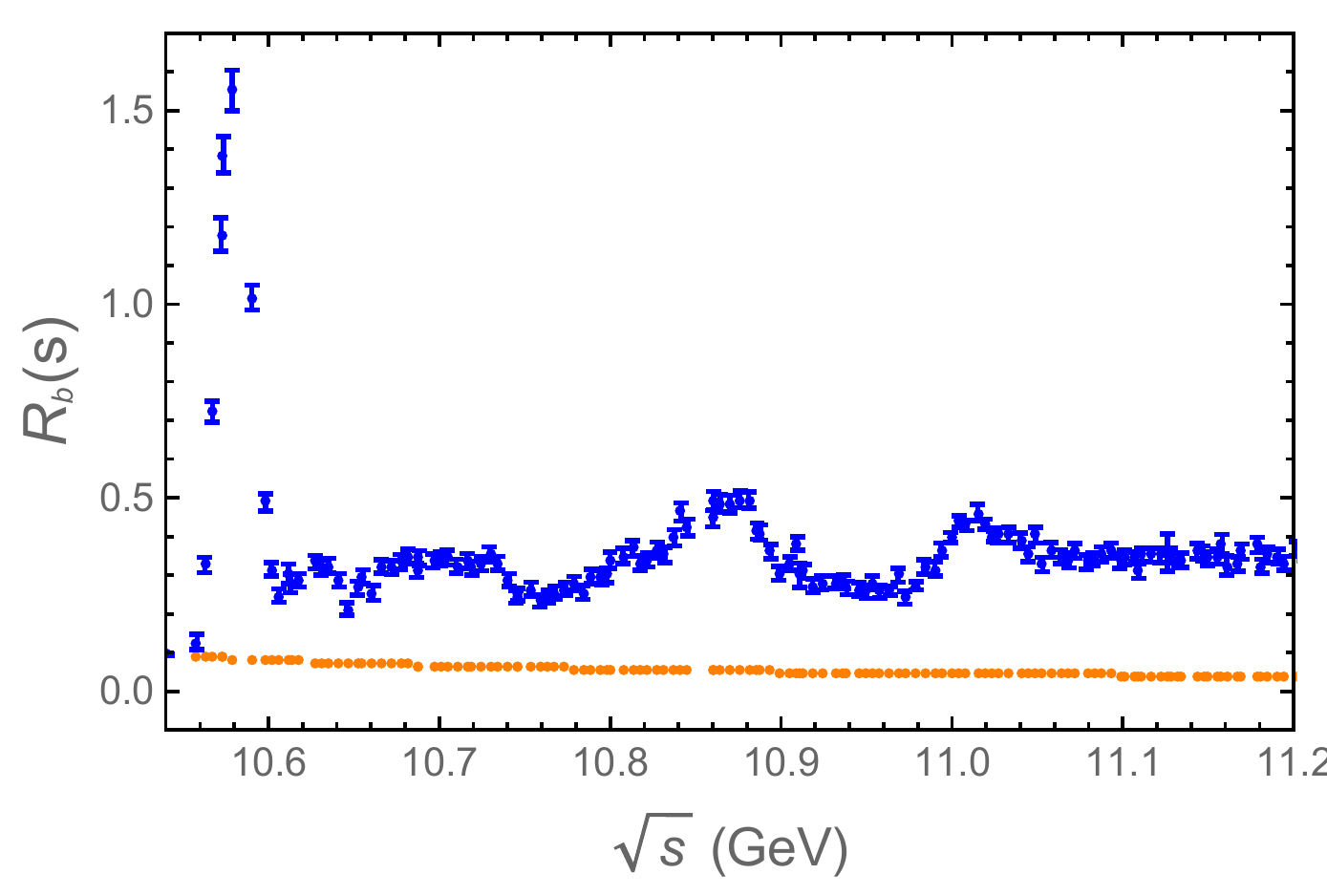}
\caption{Data for $R_b(s)$ (blue points) from the  BaBar Collaboration. 
The orange points show the initial-state radiative tail of the first three narrow states below threshold. 
Both $R_b$ data and ISR tail are taken from Ref.~\cite{Aubert:2008ab}.}
\label{RawBabarData}
\end{center}
\end{figure}

\begin{figure}[t!]
\begin{center}
\includegraphics[width=0.75\linewidth]{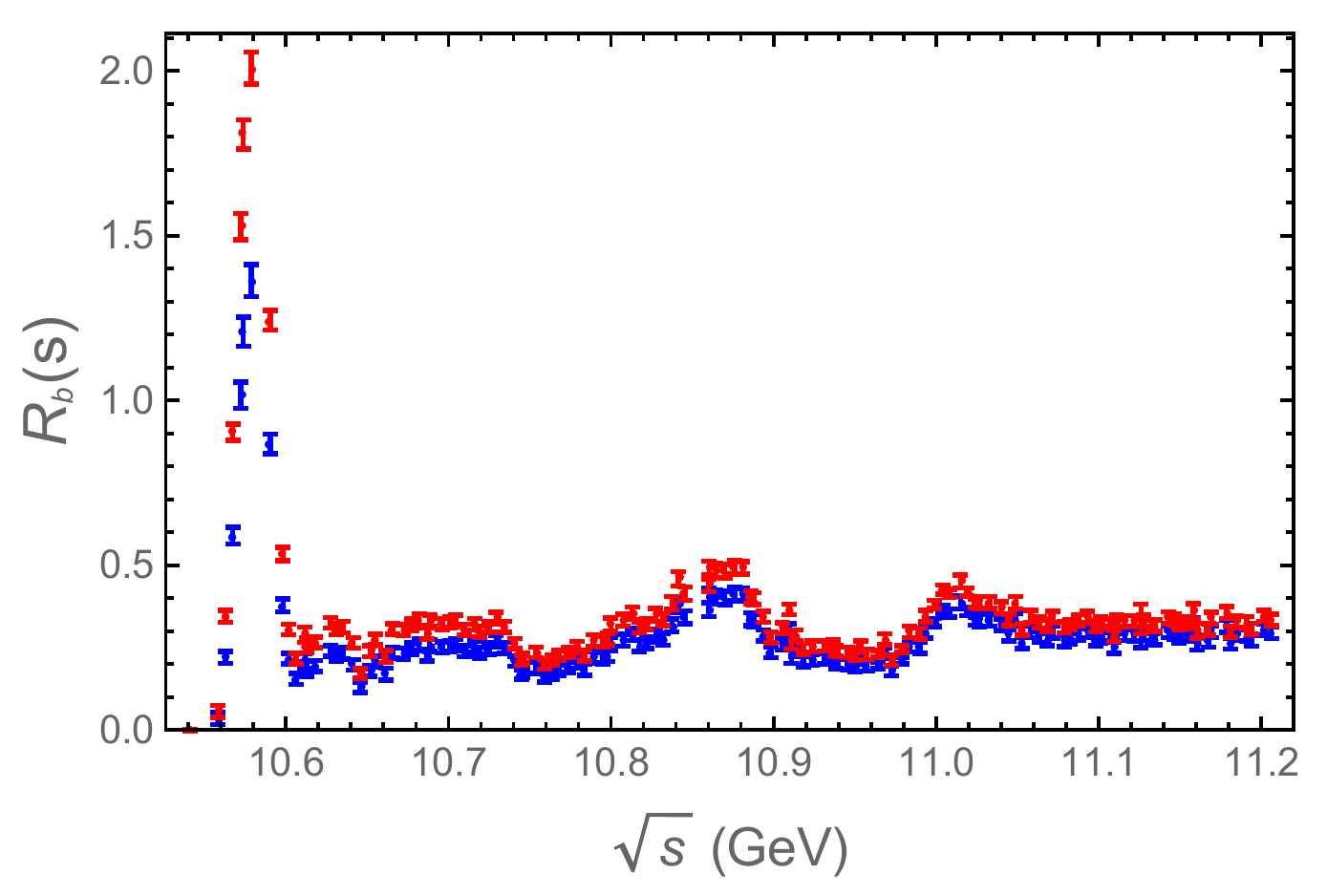}
\caption{
 BaBar data for $R_b(s)$ corrected for vacuum polarization, radiative tails and ISR (red points). 
The blue points are the same uncorrected data as shown in Fig.~\ref{RawBabarData}.}
\label{fig:ISRcomp}
\end{center}
\end{figure}

To remove initial-state radiative (ISR) effects from the continuum data after subtracting radiative tails from the resonances, 
we use the prescription following Refs.~\cite{Jadach:1988gb,Chetyrkin:1996tz} (see also Refs.~\cite{Chetyrkin:2009fv,Dehnadi:2015fra}). 
The measured $R$ ratio, $\hat R$, is given by a convolution, 
\begin{equation}
\hat{R}(s) = \int_{z_0}^1 \frac{{\rm d} z}{z} G(z,s)R(z s), 
\label{eq:rhat}
\end{equation}
of the true $R$ ratio with the radiator function $G(z,s)$ describing QED corrections. 
$G(z,s)$ is taken from Ref.~\cite{Chetyrkin:2009fv} and includes next-to-next-to-leading order contributions. 
The lower integration limit of the integral in Eq.~(\ref{eq:rhat}) should start at the onset of the continuum region, 
which we fix at $z_0 = s_0/s$ with $s_0 = (10.54~\mbox{GeV})^2$. 
The true $R$ ratio must be determined by inverting ({\em i.e.}\ unfolding) Eq.~(\ref{eq:rhat}).  
This can be done iteratively imposing the boundary condition $R(s_0) = 0$. 
This condition is automatically satisfied by the  BaBar data after subtraction of the radiative tails. 
The  BaBar data corrected for vacuum polarization, radiative tails and ISR is shown in Fig.~\ref{fig:ISRcomp} (red points). 
We also show the uncorrected data (blue points), which are the same as shown in Fig.~\ref{RawBabarData}.

 BaBar data contain an outlier at $\sqrt{s} = 10.86$~GeV (not shown in Fig.~\ref{fig:ISRcomp}). 
At this energy, there are two different experimental measurements, separated by only $\Delta \sqrt{s} = 0.0005$~GeV 
which disagree among themselves. 
Instead of removing this point, as has been suggested in Ref.~\cite{Dehnadi:2015fra}, we take the average of the two points and ascribe, 
as an error, the difference of the two measured $R$ values. 
We have checked that either option, removing the outlier or averaging with the close-by point, 
translates into a tiny difference for the experimental moments. 

\subsection{Numerical results for moments}
Experimental moments are calculated as numerical integrals over the ISR corrected $R$ values, using the trapezoidal rule. 
We collect our results in Table~\ref{tab:Data1}. 
The experimental moments ${\cal M}_n^{\rm exp}$ are affected by statistical and systematic uncertainties, 
propagated from the corresponding data errors, and we take into account correlated and uncorrelated systematic uncertainties 
following the prescription given by the  BaBar Collaboration~\cite{Aubert:2008ab}. 
For comparison, we show in Table~\ref{tab:Data1} also the moments calculated from $R_b(s)$, but using the value $\lambda_3^b = 1.53$. 
This value was obtained in our preferred scenario where the $\Upsilon(4S)$ and $\Upsilon(5S)$ resonances are included 
on top of the continuum and using the pair of moments $({\cal M}_0, {\cal M}_7)$. 
In the last column of Table~\ref{tab:Data1} we also show moments calculated from uncorrected data. 
One can see that ISR corrections are indeed very small and do not introduce an additional source of uncertainties. 

\begin{table}[t!]
\begin{center}
\begin{tabular}{|c|lll|l|}
\hline \rule[-3mm]{0mm}{9mm}
$n$ & ${\cal M}_n^{\rm exp}$ & $\lambda^{b, {\rm exp}}_3 = 0.82(20)$ & $\lambda^b_3 = 1.53$ & ${\cal M}_n^{\rm exp, no\, corr.}$ \\ 
\hline \rule[-2mm]{0mm}{7mm} 0 & 0.446(2)(11) & 0.446(11) & 0.487  & 0.453(12)\\ \rule[-2mm]{0mm}{6mm}
1 & 0.380(2)(9)  & 0.381(9)  &  0.416 & 0.384(10)\\ \rule[-2mm]{0mm}{6mm}
2 & 0.324(1)(8)  & 0.327(8)  &  0.355 & 0.328(9)\\ \rule[-2mm]{0mm}{6mm}
3 & 0.277(1)(7)  & 0.280(7)  &  0.304 & 0.279(7)\\ \rule[-2mm]{0mm}{6mm}
4 & 0.237(1)(6)  & 0.240(6)  &  0.261 & 0.238(6)\\ \rule[-2mm]{0mm}{6mm}
5 & 0.203(1)(5)  & 0.207(5)   &  0.224 & 0.204(5)\\ \rule[-2mm]{0mm}{6mm}
6 & 0.174(1)(4)  & 0.178(4)   &  0.192 & 0.174(5)\\ \rule[-2mm]{0mm}{6mm}
7 & 0.149(1)(4)  & 0.153(3)   &  0.165 & 0.149(4)\\ \rule[-2mm]{0mm}{6mm}
8 & 0.128(1)(3)  & 0.132(3)   &  0.142 & 0.128(3)\\ \rule[-2mm]{0mm}{6mm}
9 & 0.111(0)(3)  & 0.114(2)   &  0.123 & 0.110(3)\\ \rule[-2mm]{0mm}{6mm}
10 & 0.095(0)(2)  & 0.099(2)   &  0.106 & 0.094(2)\\
\hline
\end{tabular}
\end{center}
\caption{Contributions to the moments ($\times 10^{2n+1}\, \mbox{GeV}^{4n+2}$) from the restricted energy range 
$2M_B \leq \sqrt{s} \leq 11.2~\mbox{GeV}$. 
The column labeled ${\cal M}_n^{\rm exp}$ is obtained by direct integration over corrected data. 
The first error is due to the uncorrelated statistical and systematic uncertainties of the data, while the second is the correlated one. 
The third and fourth columns show the moments calculated from our {\em ansatz\/} for $R_b(s)$ with 
$\lambda^{b, {\rm exp} }_3 = 0.82(20)$ (column 3) and $\lambda^b_3 = 1.53$ (column 4) as input. 
In both cases, $\hat{m}_b=4.1802$~GeV was used. 
The last column collects the experimental moments when  BaBar data is used without any kind of correction or subtraction.}
\label{tab:Data1}
\end{table}

In Table~\ref{tab:expm} we compare our determination of moments with those from Ref.~\cite{Dehnadi:2015fra} and Ref.~\cite{Chetyrkin:2009fv}. 
To do so, we have to adjust the energy range correspondingly. 
For both references the lower limit of the energy range was chosen at $\sqrt{s} = 10.62~\mbox{GeV}$. 
The upper integration limit was $\sqrt{s} = 11.20~\mbox{GeV}$ in Ref.~\cite{Dehnadi:2015fra} and 
$\sqrt{s} = 11.24~\mbox{GeV}$ in Ref.~\cite{Chetyrkin:2009fv}. 
We also follow Refs.~\cite{Dehnadi:2015fra,Chetyrkin:2009fv} and subtract the $\Upsilon(4S)$ resonance, 
which is parameterized by a Breit-Wigner distribution, as well as its radiative tail. 
Above $\sqrt{s} = 11.20~\mbox{GeV}$, we use our {\em ansatz\/} to extrapolate up to $11.24$~GeV. 
As can be seen from Table~\ref{tab:expm}, we find good agreement with both references. 

\begin{table}[t!]
\begin{center}
\begin{tabular}{|l|ll||ll|}
\hline \rule[-3mm]{0mm}{9mm}
$n$ & Ref.~\cite{Dehnadi:2015fra} & This work & Ref.~\cite{Chetyrkin:2009fv} & This work \\  
\hline \rule[-2mm]{0mm}{7mm}
0 & $-$ & 0.321(12) & $-$ & 0.336(12) \\ \rule[-2mm]{0mm}{6mm}
1 & 0.270(2)(9) & 0.269(10) & 0.287(12) & 0.281(10) \\ \rule[-2mm]{0mm}{6mm}
2 & 0.226(1)(8) & 0.226(9) & 0.240(10) & 0.235(9) \\ \rule[-2mm]{0mm}{6mm}
3 & 0.190(1)(7) & 0.189(8) & 0.200(8) & 0.197(8) \\ \rule[-2mm]{0mm}{6mm}
4 & 0.159(1)(6) & 0.159(7) & 0.168(7) & 0.165(7) \\ \rule[-2mm]{0mm}{6mm}
5 & $-$  & 0.133(6) & $-$ & 0.138(6) \\ \rule[-2mm]{0mm}{6mm}
6 & $-$ & 0.112(5) & $-$ & 0.116(5) \\ \rule[-2mm]{0mm}{6mm}
7 & $-$ & 0.094(4) & $-$ & 0.097(4) \\ \hline
\end{tabular}
\end{center}
\caption{Comparison of moments with Ref.~\cite{Dehnadi:2015fra} (left section) and Ref.~\cite{Chetyrkin:2009fv} (right section). 
In the first case, moments ${\cal M}_n$ are calculated from data in the range 
$10.62~\mbox{GeV} \leq \sqrt{s} \leq 11.20~\mbox{GeV}$, while in the second case the energy range is 
$10.62~\mbox{GeV} \leq \sqrt{s} \leq 11.24~\mbox{GeV}$. 
Our calculation uses experimental data up to $\sqrt{s} = 11.20$~GeV and an extrapolation based on our {\em ansatz\/} 
to cover the energy range up to $\sqrt{s} = 11.24$~GeV. 
In both cases, the $\Upsilon(4S)$ resonance including its radiative tail is subtracted.}
\label{tab:expm}
\end{table}

\subsection{Determination of $\lambda_3^{b, {\rm exp}}$}
Now that the experimental moments are determined, we proceed to calculate $\lambda_3^{b,{\rm exp}}$ by solving the equation,
\be 
\int_{(2 M_B)^2}^{(11.20{\rm \, GeV})^2} \frac{{\rm d} s}{s} R_b^{\rm cont}(s) = {\cal M}_0^{\rm Data} = 0.446 \pm 0.011,
\label{eq:expmoment}
\ee
where $R_b^{\rm cont}(s) $ is defined in Eq.~(\ref{eq:ansatz}). 
The $b$ quark mass $\hat{m}_b(\hat{m}_b)$ is fixed in Eq.~(\ref{eq:expmoment}) to the value 
obtained from a selected pair of moments $({\cal M}_0, {\cal M}_n)$ as described in the previous section. 
The solution of Eq.~(\ref{eq:expmoment}) is called $\lambda_3^{b, \rm exp}$. 
Results are shown in Table~\ref{tab:MomentsBudget4S5S} already discussed above. 
In each case, the value obtained for $\lambda_3^{b, \rm exp}$ 
is compared with $\lambda_3^b$ determined from the corresponding pair of sum rules. 
For the default case where we add the $\Upsilon(4S)$ and $\Upsilon(5S)$ resonances on top of the continuum 
and use the $0^{\rm th}$ and $7^{\rm th}$ moments, we find $\lambda_3^{b, \rm exp} = 0.82\pm 20$. 
The difference between this value and the one determined from the pair of moments 
($\lambda_3^b = 1.53$, see Table~\ref{tab:MomentsBudget4S5S}) 
corresponds to a difference in terms of the $b$ quark mass of $3.5 \pm 1.0$~MeV. 

We have used the  BaBar data since it covers an energy range large enough to extract a reliable description of the continuum region. 
The Belle Collaboration~\cite{Santel:2015qga} also provides a measurement of $R_b(s)$, 
but only the narrow energy range between $\sqrt{s} = 10.620$ and 11.047~GeV is covered 
with the first three experimental points quite disconnected from the fine-scan around the $\Upsilon(5S)$ and $\Upsilon(6S)$ resonances, 
{\em i.e.}\ $10.754~\mbox{GeV} \leq \sqrt{s} \leq 11.047~\mbox{GeV}$. 
If we use Belle data we find that this short energy range contributes $0.198(7)$ to the $0^{th}$ experimental moment, 
to be compared with $0.172(5)$ from  BaBar data for the same energy region. 
These results are compatible at the $3 \sigma$ level only. 
Such a difference could by attributed to the different treatment of QED radiative effects of the narrow resonances in the case of Belle data. 
For our calculation of the $0^{th}$ moment we have used Belle data corrected for vacuum polarisation effects, 
but without subtracting radiative tails. 
The Belle Collaboration does not provide the corresponding information. 
If we had used the radiative tail provided by  BaBar, we would find $0.167(7)$ for the $0^{th}$ moment. 
This would bring the values of the $0^{th}$ moment calculated from Belle or from  BaBar data in very good agreement. 

\begin{table}[t]
\begin{center}
\begin{tabular}{|c|cc|}
\hline \rule[-3mm]{0mm}{9mm}
$n$ & Data $+$ pQCD  & Data + continuum {\em ansatz\/} \\ 
\hline \rule[-2mm]{0mm}{7mm}
0 & 2.532 (11) &2.165(69)  \\ \rule[-2mm]{0mm}{6mm}
1 & 1.637 (9) & 1.401(42)  \\ \rule[-2mm]{0mm}{6mm}
2 & 1.103 (8) &0.947(26)  \\ \rule[-2mm]{0mm}{6mm}
3 & 0.773 (7) &0.667(17)  \\ \rule[-2mm]{0mm}{6mm}
4 & 0.560 (6) &0.487(11)  \\ \rule[-2mm]{0mm}{6mm}
5 & 0.418 (5) &0.368(8)  \\ \rule[-2mm]{0mm}{6mm}
6 & 0.321 (4) &0.284(5)  \\ \rule[-2mm]{0mm}{6mm}
7 & 0.251 (4) &0.224(4)  \\ \rule[-2mm]{0mm}{6mm}
8 & 0.200 (3) &0.181(3)  \\ \rule[-2mm]{0mm}{6mm}
9 & 0.161 (3) &0.147(2)  \\ \rule[-2mm]{0mm}{6mm}
10 & 0.132 (2) &0.122(1) \\
\hline
\end{tabular}
\end{center}
\caption{Predictions for the moments in the region $2 M_B \leq \sqrt{s} \leq 15.0~\mbox{GeV}$ from data and an extrapolation using pQCD 
including the known heavy quark mass corrections (2nd column) or 
our {\em ansatz\/} with $\hat{m}_b(\hat{m}_b) = 4.1802$~GeV and $\lambda_3^{b,exp}= 0.82(20)$ (3rd column). 
All numbers are given in units of $10^{- (2n+1)}~{\rm GeV}^{-2n}$. 
Errors in the 2nd column are from experimental moments only as no uncertainty is assigned to the contribution from pQCD. 
The errors in the 3rd column combine those from the experimental moments and from $\Delta \lambda_3^b = \pm 0.20$.}
\label{tab:expm2}
\end{table}

In our previous analysis of data for charm quark production \cite{Erler:2016atg} we found that using data in an energy range 
between 3.7 and 5~GeV, extended by using pQCD above, can lead to a consistent picture and a reliable determination of the charm quark mass. 
A correspondingly large energy window for the bottom quark would cover energies up to 15~GeV, 
{\em i.e.}\ roughly one unit of the heavy quark mass above threshold. 
Unfortunately, data are available only for $\sqrt{s} \leq 11.2$~GeV. 
The treatment of the energy range $11.2~\mbox{GeV} \leq \sqrt{s} \lesssim 15~\mbox{GeV}$ 
requires special care and may lead to additional uncertainties. 
In Ref.~\cite{Chetyrkin:2009fv} it was argued that the gap above $\sqrt{s} = 11.2$~GeV should be described by pQCD. 
However, this introduces a discontinuity with the experimental data. 
In Ref.~\cite{Dehnadi:2015fra} a smooth polynomial fit was used instead. 
We opt for using our own {\em ansatz\/}, which approaches pQCD only for $s\to \infty$. 
In Table~\ref{tab:expm2} we compare two possible options: 
calculating moments from data in the window $2 M_B \leq \sqrt{s} \leq 11.2~\mbox{GeV}$, combined either with pQCD or 
our {\em ansatz\/} for $R_b(s)$, both up to $15~\mbox{GeV}$. 
If one uses data and pQCD above $\sqrt{s} = 11.2$~GeV, {\em i.e.}\ a prescription with a discontinuity, 
one obtains moments which are larger than for the case where we use our {\em ansatz\/} with a smooth $\sqrt{s}$-dependence.  
Correspondingly, this will result in smaller values for the bottom mass, as found in Ref.~\cite{Chetyrkin:2009fv}. 

Differences in the $b$ quark mass determination between Refs.~\cite{Chetyrkin:2009fv,Dehnadi:2015fra} and our work 
can thus be traced to a different prescription for including contributions to the moments from an energy range where no data are available. 
One could argue that this should be considered as an additional systematic error for the $b$ quark mass. 
Experimental data covering the energy range between $11$ and $15$~GeV are definitely needed to ultimately solve this issue. 
Until such data will become available, we believe that a description of the unknown part of $R_b(s)$ 
with a smooth function is preferable over one with a discontinuity.

\section{Conclusions}
\label{sec:conclusions}
\begin{figure}[t]
\begin{center}
\includegraphics[width=0.94\textwidth]{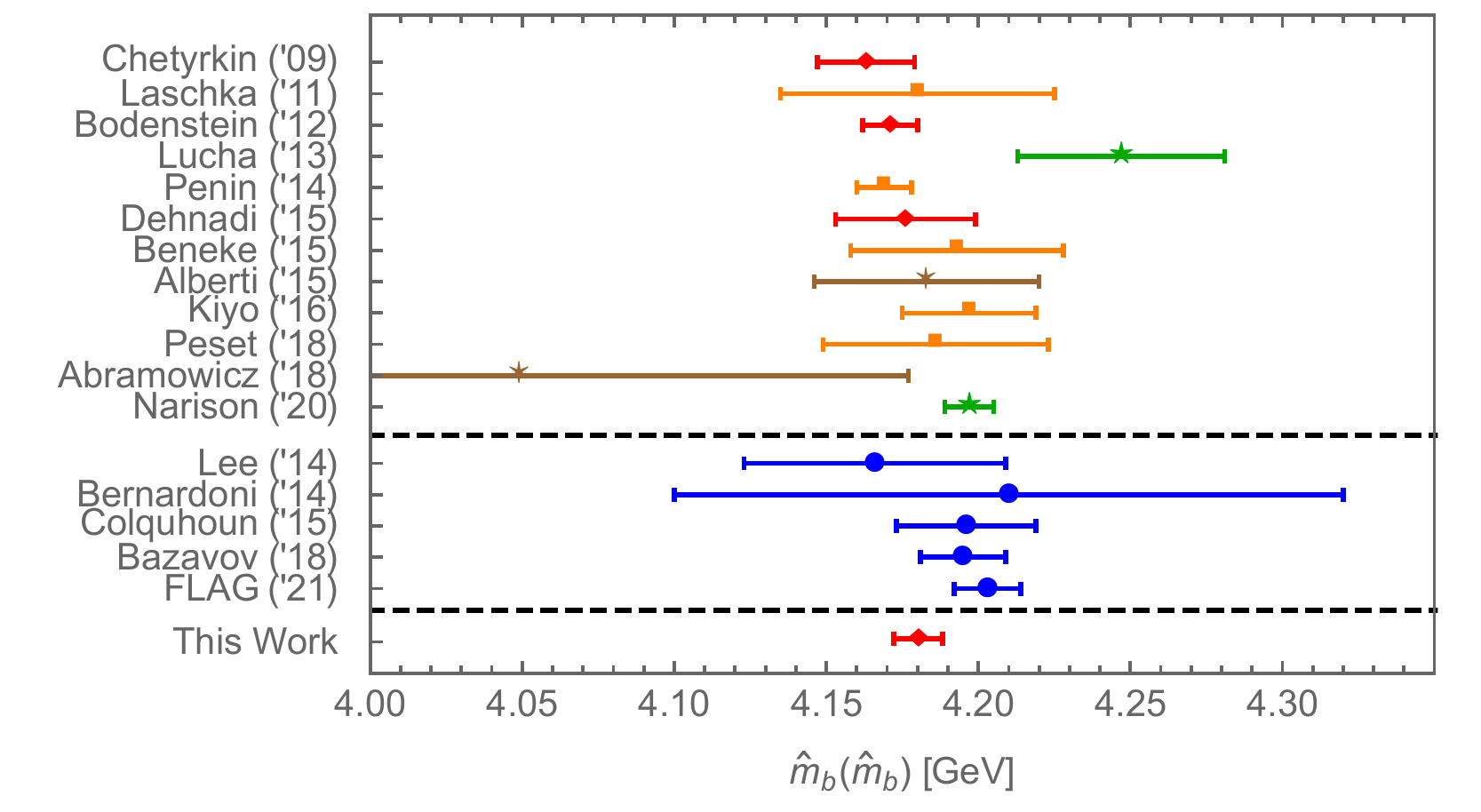}
\caption{Recent bottom quark mass determinations from phenomenological studies (upper part; red, orange, brown and green symbols) 
and lattice calculations (lower part; blue points). 
See text for details and references.}
\label{fig:comparison}
\end{center}
\end{figure}

Our final result, $\hat m_b(\hat m_b) = (4180.2 - 108.5 \Delta \hat{\alpha_s} \pm 7.9)~{\rm MeV}$ with $\Delta \hat \alpha_s = \hat \alpha_s(M_Z) - 0.1182$, is in good agreement with other determinations of the bottom quark mass that can be found in the literature. 
We show a comparison in Figure~\ref{fig:comparison} where we group the results in two sets and in chronological order within each set. 

The first set is based on phenomenological approaches, extracting $\hat{m}_b(\hat{m}_b)$ by comparing theory predictions with data. 
This includes other results based on relativistic sum rules, 
Chetyrkin~(2009)~\cite{Chetyrkin:2009fv}, 
Bodenstein~(2012)~\cite{Bodenstein:2011fv}, and
Dehnadi~(2015)~\cite{Dehnadi:2015fra}, shown as red diamonds.
The methodology of these publications is closest to our own approach. 
Our higher value for $\hat{m}_b(\hat{m}_b)$ can be traced to the treatment of the intermediate energy behavior where our method approaches 
the perturbative regime of QCD at higher energies, as discussed in detail above. 
We also display results based on non-relativstic sum rules (orange squares),
Laschka~(2011)~\cite{Laschka:2011zr}, 
Penin~(2014)~\cite{Penin:2014zaa}, 
Beneke~(2015)~\cite{Beneke:2014pta}, 
Kiyo~(2016)~\cite{Kiyo:2015ufa}, and
Peset~(2018)~\cite{Peset:2018ria}, 
as well as on other sum rule methods (green stars),
Lucha~(2013)~\cite{Lucha:2013gta} and
Narison~(2020)~\cite{Narison:2019tym}. 
The bottom quark mass (brown stars) was also determined as a by-product in a global fit 
to inclusive semileptonic $B$-meson decays to obtain the CKM matrix element $V_{cb}$, 
Alberti~(2015)~\cite{Alberti:2014yda},
and from an analysis of deep inelastic scattering data at HERA compared with perturbative QCD calculations, 
Abramowicz~(2018)~\cite{H1:2018flt}. 

The results in the lower part of Fig.~\ref{fig:comparison} (blue points) are lattice QCD calculations. 
They are based on an improved non-relativistic QCD action, 
Lee~(2014)~\cite{Lee:2013mla}, 
on Heavy Quark Effective Theory non-perturbatively matched to QCD,
Bernardoni~(2014)~\cite{Bernardoni:2013xba}, 
on using time-moments of the vector current-current correlator,
Colquhoun~(2015)~\cite{Colquhoun:2014ica}, 
as well as the MILC highly improved staggered quark ensembles with four flavors of dynamical quarks, 
Bazavov~(2018)~\cite{FermilabLattice:2018est}. 
We also show the average of the 2021 FLAG Review~\cite{Aoki:2021kgd} for $N_f=2+1+1$.


\section*{Acknowledgments}
This work was supported by the German-Mexican research collaboration grant No.\ 278017 (CONACyT) and No.\ SP 778/4-1 (DFG).
P.M.\ has received support from the Secretaria d'Universitats i Recerca del Departament d'Empresa i Coneixement 
de la Generalitat de Catalunya under the grant 2017SGR1069, 
the Ministerio de Economia, Industria y Competitividad (grant FPA2017-86989-P and SEV-2016-0588), and 
the Ministerio de Ciencia e Innovaci\'on (grant PID2020-112965GB-I00).

\end{document}